\documentclass[letterpaper]{article} 
\usepackage{aaai23}  
\usepackage{times}  
\usepackage{helvet}  
\usepackage{courier}  
\usepackage[hyphens]{url}  
\usepackage{graphicx} 
\usepackage{natbib}  
\usepackage{caption} 
\usepackage{algorithm}
\usepackage{algorithmic}
\usepackage{amsmath} 
 
\usepackage{bm}
\usepackage{threeparttable}
\usepackage{booktabs} 
\usepackage{multirow}
\usepackage{enumitem}
\usepackage{amssymb}
\usepackage{pdfrender}
\usepackage{pifont}
\usepackage{soul}
\usepackage[hang,flushmargin]{footmisc}

\usepackage{tabularx}
\usepackage{diagbox}

\usepackage{todonotes}
\usepackage{comment}

\usepackage{newfloat}
\usepackage{listings}
\DeclareCaptionStyle{ruled}{labelfont=normalfont,labelsep=colon,strut=off} 
\floatstyle{ruled}
\newfloat{listing}{tb}{lst}{}
\floatname{listing}{Listing}
\title{Structure Aware Incremental Learning with Personalized Imitation Weights for Recommender Systems}
\author {
    Yuening Wang\textsuperscript{\rm 1},
    Yingxue Zhang\textsuperscript{\rm 1},
    Antonios Valkanas\textsuperscript{\rm 2}\thanks{Work done as an intern at Huawei Noah's Ark Lab},
    Ruiming Tang\textsuperscript{\rm 1},
    Chen Ma\textsuperscript{\rm 3},
    Jianye Hao\textsuperscript{\rm 1, \rm 4},
    Mark Coates\textsuperscript{\rm 2}
}
\affiliations {
    \textsuperscript{\rm 1} Huawei Noah's Ark Lab,
    \textsuperscript{\rm 2} McGill University,
    \textsuperscript{\rm 3} City University of Hong Kong,
    \textsuperscript{\rm 4} Tianjin University,
    yuening.wang@huawei.com, yingxue.zhang@huawei.com, antonios.valkanas@mail.mcgill.ca, tangruiming@huawei.com, chenma@cityu.edu.hk, haojianye@huawei.com, mark.coates@mcgill.ca
}

\usepackage{bibentry}

\begin{document}

\maketitle

\begin{abstract}
Recommender systems now consume large-scale data and play a significant role in improving user experience. Graph Neural Networks (GNNs) have emerged as one of the most effective recommender system models because they model the rich relational information. The ever-growing volume of data can make training GNNs prohibitively expensive. To address this, previous attempts propose to train the GNN models incrementally as new data blocks arrive. 
Feature and structure knowledge distillation techniques have been explored to allow the GNN model to train in a fast incremental fashion while alleviating the catastrophic forgetting problem. 
However, preserving the same amount of the historical information for all users is sub-optimal since it fails to take into account the dynamics of each user's change of preferences. 
For the users whose interests shift substantially, retaining too much of the old knowledge can overly constrain the model, preventing it from quickly adapting to the users’ novel interests. 
In contrast, for users who have static preferences, model performance can benefit greatly from preserving as much of the user's long-term preferences as possible.
In this work, we propose a novel training strategy that adaptively learns personalized imitation weights for each user to balance the contribution from the recent data and the amount of knowledge to be distilled from previous time periods.
We demonstrate the effectiveness of learning imitation weights via a comparison on five diverse datasets for three state-of-art structure distillation based recommender systems. The performance shows consistent improvement over competitive incremental learning techniques.
\end{abstract}

\section{Introduction}

The growth of online services has rendered recommender systems a vital part of providing personalized recommendations to users. Making highly relevant recommendations improves user experience and increases the service provider's revenue. Deep learning models are becoming more prevalent in all aspects of recommender system design due to their superiority in constructing high-quality user and item representations in an end-to-end fashion~\cite{covington2016deep,deepfm,cheng2016wide}.  There is a recent trend to formulate the recommendation problem as a learning task on graphs because of the rich relational information that graphs can model. Much of the data from recommender systems can naturally be expressed using graph structures~\cite{gcmc_vdberg2018,NGCF_wang19,wang2021graph,sun2020_mgcf}. For example, we can use the user-item bipartite interaction graph, an item similarity graph, a user-user graph derived from social network exchanges, and an additional knowledge graph to improve the representation learning process. Graph Neural Network (GNN) based recommender systems have emerged as one of the most effective models because the message-passing paradigm allows sufficient modeling of the relational information in the data. However, training GNNs on large-scale graphs can be prohibitively expensive~\cite{ying2018,zou2019layer,chiang2019cluster,graphsaint-iclr20,gag_streaming_gnn,xu2020graphsail}, which makes deploying models with GNN backbone networks extremely challenging on large-scale recommender systems, especially since there is a need to satisfy a strict time constraint for online systems.


One approach to address the computation issue is to train the deep learning models incrementally as new data blocks arrive~\cite{kirkpatrick2017overcoming, Shmelkov_2017_ICCV, Castro2018E2E, rebuffi2017icarl, mallya2018packnet_IL, xu2018reinforced_IL, gag_streaming_gnn}. However, directly using the data from the incremental block to fine-tune a model can lead to catastrophic forgetting~\cite{kirkpatrick2017overcoming, Shmelkov_2017_ICCV}. 
Because of their superiority in terms of efficiency and performance, knowledge distillation approaches~\cite{Castro2018E2E, kirkpatrick2017overcoming,xu2020graphsail,Wang2021GraphSA} are preferable for models with a GNN backbone architecture. Benefiting from the knowledge distillation paradigm, key information from the historical data is preserved and transferred to the student model trained using only newly arrived data. This is achieved by regularizing the distance between the representations of the teacher and the student models. Both feature and structure knowledge distillation techniques have been explored; these allow the GNN model to better preserve both the node feature and structure information in the previous training data while enjoying the fast incremental training process~\cite{yang2020distilling,xu2020graphsail,Wang2021GraphSA}.

\begin{figure}
    \centering
    \includegraphics[width=0.9\columnwidth]{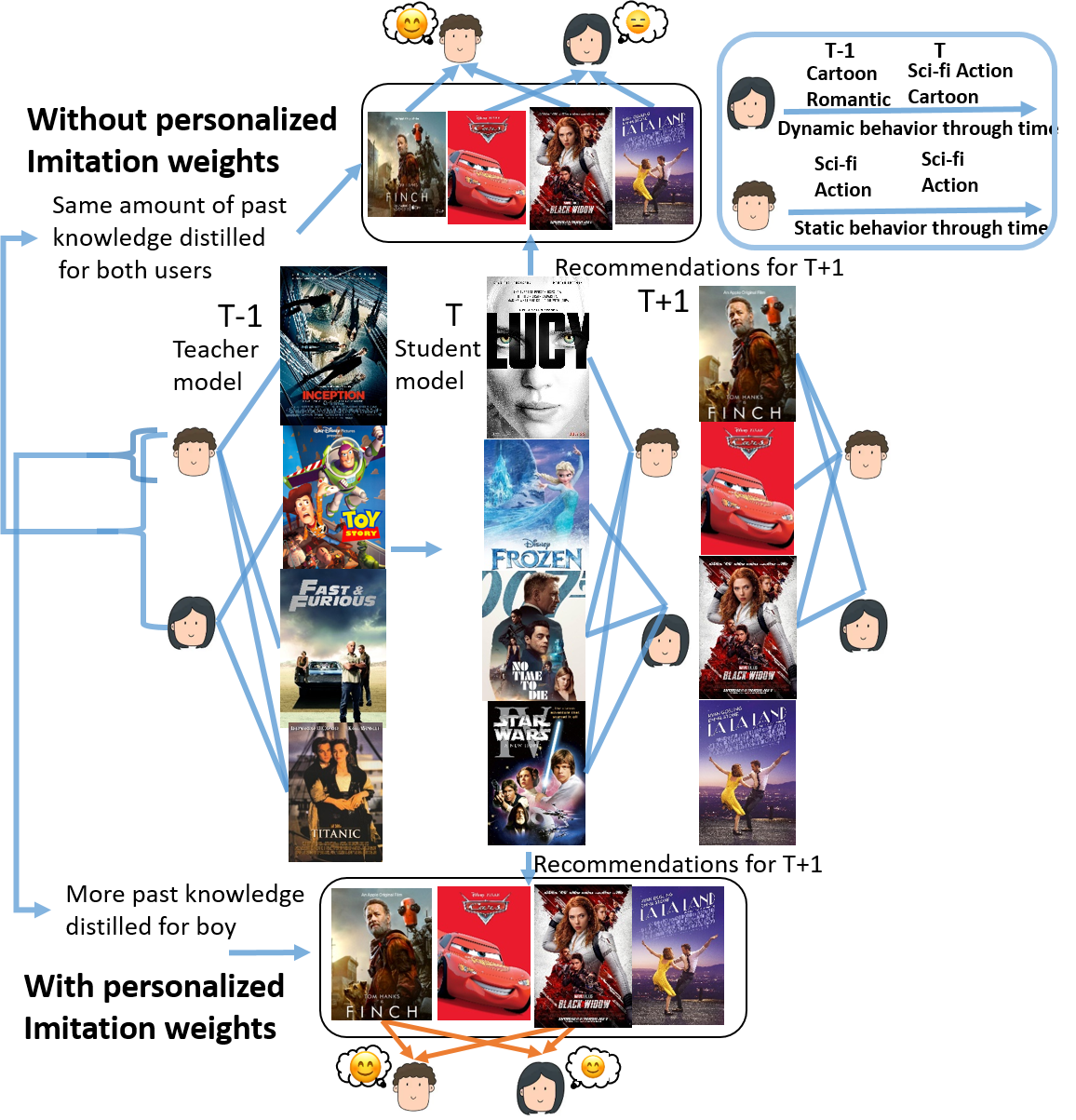}
    \caption{Illustration of the motivation of the designed framework where the boy has a constant preference from time $t-1$ to $t$, while the girl's interests change more dynamically. Capturing the interest shift difference between users in the form of a personalized incremental learning scheme will be beneficial.}
    \label{fig:demo}
    \vspace{-0.2cm}
\end{figure}

However, preserving the same amount of historical information for all users without any distinction between them might be sub-optimal since it fails to take into account the dynamics of each user's potential change of preferences. For users whose interests shift substantially, retaining too much of the old knowledge from the past via the knowledge distillation process might prevent the model from quickly adapting to the users’ latest interests. In contrast, for users who have more static preferences, model performance can benefit greatly from preserving as much of the user's long-term preference as possible.  We illustrate this with an example in Figure~\ref{fig:demo}.

Thus, how to design an incremental learning training scheme that can model the dynamics of users' personalized preference change to determine how much knowledge to preserve from the past is an intriguing and important research question. We address this research question by targeting an important hyper-parameter in knowledge distillation objective functions, named the \textit{imitation weight}, which is used to balance the contributions to the overall loss from the new data  and from distillation.


In this work, following the above intuition, we propose a novel end-to-end training strategy that adaptively learns \textit{personalized imitation weights} for each user to better balance the contributions from the recent data and the amount of knowledge to be distilled from previous time windows. Specifically, we first model each user’s preference as a distribution over the distance to item cluster centers, with the clusters being obtained by a deep structural clustering method applied to the user-item bipartite graph. The cluster learning process is integrated into the overall training procedure. Then we construct, for each user, a state vector that encodes the distance between two preference distributions associated with that user, which are derived from consecutive training blocks. This state vector is passed as the input to a weight generator parameterized by a neural network to produce a user-specific imitation weight. This personalized imitation weight determines how much information pertinent to the user is inherited from the teacher (historical) model. Our proposed approach is not restricted to a specific backbone architecture or incremental learning procedure. It can easily be integrated with multiple existing state-of-the-art methods.

To summarize, the main contributions of the paper are:
\begin{enumerate}

    \item We demonstrate that explicitly assessing the user interest shift between consecutive training blocks and using this signal to learn a user-specific imitation weight is an important modeling factor. It can significantly impact the performance of knowledge distillation-based incremental learning techniques. 
    To the best of our knowledge, this is the first incremental learning training scheme that explicitly models user change of preferences. 
    
    \item We propose a novel training strategy that adaptively learns personalized imitation weights for each user to balance the contribution from recent data and the amount of knowledge that is distilled from previous periods. 
    
    \item We demonstrate the effectiveness of learning imitation weights via a thorough comparison on five diverse datasets. Our best-performing model improves the SOTA method by 2.30\%. We integrate our proposed training procedure with three recent SOTA incremental learning techniques for recommender systems. We show consistent improvement over the non-adaptive counterparts.

\end{enumerate}
\section{Related Work}



\subsection{Incremental learning}

Incremental learning is a branch of machine learning that aims to develop models which are updated continuously with new data. However, naively training on new batches of data as they arrive leads to the problem of \emph{catastrophic forgetting} that the model forgets previously learned information and is overly biased to new data~\cite{kirkpatrick2017overcoming, Shmelkov_2017_ICCV, Castro2018E2E}. 

There are two main groups of approaches to combat this issue:(i) regularization-based knowledge distillation~\cite{Castro2018E2E, kirkpatrick2017overcoming,xu2020graphsail,Wang2021GraphSA} and (ii) experience replay, also referred to as reservoir sampling~\cite{prabhu2020gdumb, Ahrabian2021StructureAE}. Reservoir methods sample a data reservoir containing the most representative historical data and replay it while learning new tasks to alleviate forgetting. Some key reservoir sampling works such as iCarl~\cite{rebuffi2017icarl} and GDumb~\cite{prabhu2020gdumb} focus on optimizing the reservoir construction either via direct optimization or via greedy heuristics. Recent work on graph recommender systems expands on the GDumb heuristic and proposes inverse degree sampling of nodes for reservoir construction~\cite{Ahrabian2021StructureAE}.

Regularization techniques typically introduce penalty parameters in the loss function to prevent the model weights from ``drifting'' too far from their tuned values from historical data blocks, thus preventing forgetting~\cite{yang2019adaptive, xu2020graphsail, Wang2021GraphSA}. Knowledge distillation is one of the most common regularization approaches. Knowledge distillation refers to the process of transferring knowledge from a large and complex teacher model to a smaller student model without significant loss in performance~\cite{hinton2015distilling}. In incremental learning, the \textit{teacher model} is trained on the historical data and the \textit{student model} is trained on the new data.

Although both incremental learning and sequential learning take advantage of historical information, sequential recommendation learning is different from incremental learning in several important aspects.  First, incremental learning aims to substantially reduce the training sample number by inheriting the knowledge from the previously trained model with knowledge distillation or experience reply. In contrast, sequential recommender systems focus on better characterizing the user's long-term or short-term interaction sequences through memory units~\cite{hidasi2018recurrent}, Recurrent Neural Networks (RNNs)or attention design~\cite{kang2018self,fan2021continuous}. The former is a training strategy in the scenario that requires incremental updates and the latter is a specific model architecture to handle the given sequence of data. Second,  incremental learning in the context of recommender systems is agnostic to any type of backbone architecture. Both sequential and non-sequential models should be compatible with the incremental training method.
\subsection{Incremental Learning on Graph Structured Data}
Graph representation learning techniques have become a mainstream tool for collaborative filtering and recommender systems~\cite{sun2020_mgcf, ying2018, NGCF_wang19, he2020lightgcn}. However, they suffer from a computation and memory burden introduced by either the neighborhood sampling process, message passing procedureor the storage of the adjacency matrix, which prevents GNN models from satisfying a strict training time constraint for online systems~\cite{xu2020graphsail}.
Several incremental recommender system designs have been proposed to tailor the GNN models to better preserve the structural information.  GraphSAIL~\cite{xu2020graphsail} employs knowledge distillation at the node level, the node neighborhood, and the global graph level.  LSP\_s~\cite{yang2020distilling} minimizes the distance between structure-related distributions drawn from the model trained at previous time steps and the fine-tuned model.
SGCT~\cite{Wang2021GraphSA} introduces a contrastive approach to knowledge distillation. The objective of SGCT is to maximize the lower bound of the mutual information between pairs of adjacent node embeddings from the student and the teacher model.
LWC-KD~\cite{Wang2021GraphSA} improves over SGCT by considering intermediate layer embedding distillation and additional contrastive distillation on the user-user and item-item graph.

\textbf{Novelty of our work} Prior works focus on universally distilling as much information as possible for all users, without distinguishing between them. However, this is not always optimal or desirable as the interests of some users may shift quickly over time and the recommendation models should be able to adapt quickly to the new user preference. Our work proposes an adaptive weight mechanism to learn the amount of knowledge to distill for each user. We show experimentally that personalizing the distillation strength by assessing how rapidly each user's interests are changing can lead to significantly better recommendation performance for state-of-art backbones.

\section{Methodology}
In this section, we present the proposed  
\textbf{\underline{S}}tructure \textbf{\underline{A}}ware \textbf{\underline{I}}ncremental \textbf{\underline{L}}earning with \textbf{\underline{P}}ersonalized \textbf{\underline{I}}mitation
 \textbf{\underline{W}}eight frameworks, abbreviated to SAIL-PIW, which exploits personalized imitation weights by characterizing the user interest distribution shift between the historical data and the newly arrived incremental data. The personalized imitation weight is used in the knowledge distillation loss to balance the contribution of the recent data and the amount of historical knowledge to be distilled from previous time periods. To better model each user’s preference, we learn a distribution over the distance to item cluster centers, with the clusters being obtained by a structural clustering method applied to the user-item bipartite graph. Once we obtain the user interest distribution, we then use the difference of the interest distribution between the incremental learning block and the teacher model as a user interest shift indicator. This shift indicator models the change of user preferences between the most recent training block and the newly arrived incremental block. The shift indicator is then passed to a weight generator, parameterized by a multi-layer perceptron to output a user-specific imitation weight. Our proposed framework is trainable in an end-to-end fashion via back-propagation. 
 
\begin{figure*}
    \centering
    \includegraphics[width=1.0\linewidth]{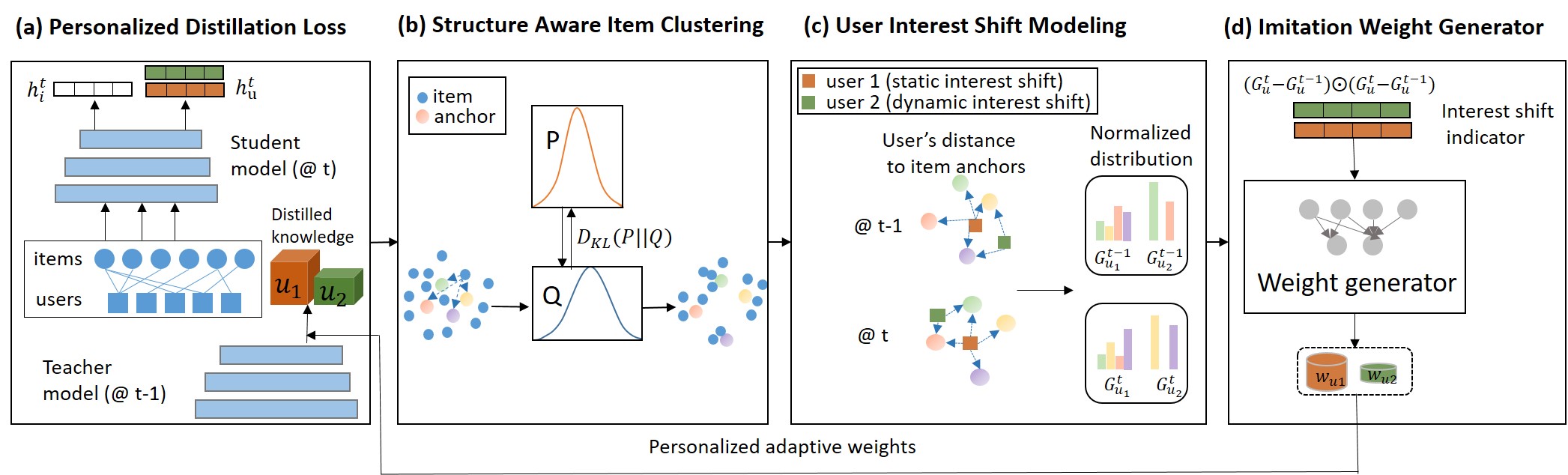}
    \caption{\textbf{The overall framework of our proposed SAIL-PIW model}. \textbf{(a)}. The student model at time t takes a new user-item interaction graph as input to learn users' and items' embeddings, regularized by the distilled knowledge from the teacher model at time point $t-1$. The knowledge distilled for static user 1 and dynamic user 2 is controlled by the weights learned in the following steps. \textbf{(b)}. We learn the distance of an item embedding from the GNN model to a learnable item anchor embedding. It minimizes the KL divergence between learned items' distribution to clusters P and target distribution Q. \textbf{(c)}. The user's distribution to item anchors is calculated for both time point $t-1$ and time point $t$. Two types of users are illustrated: user 1 with low interest shift and user 2 with high interest shift. \textbf{(d)}. The user interest shift is calculated as the difference of user's normalized distance distribution to item cluster centers. The weight generator takes the user interest shift indicator as input to produce a personalized imitation weight, which is used to control knowledge distilled for each user as mentioned in \textbf{(a)}. }
    \label{fig:model_overview}
    \vspace{-1em}
\end{figure*}

 The overall architecture of our model is presented in Figure~\ref{fig:model_overview}.
 In the following subsections, we initially describe in detail the individual components of our proposed method. Thus, we delicately separate sections into (i) the proposed personalized distillation loss, (ii) the architecture of the imitation weight generator, (iii) the structure-aware item clustering technique to obtain the cluster center embedding, and finally (iv) the way we construct the metric to characterize the user interest shift. 
 It is important to note that our framework is not limited to a particular GNN model, graph-based recommender system architectureor a specific incremental learning framework.  We illustrate promising results of using GraphSAIL, SGCT and LWC-KD  as backbones. Thus, our proposed solution is highly appealing in real-world settings as it can be readily applied on top of existing graph-based systems regardless of the base models that these systems already employ.

\subsection{Personalized Distillation Loss}
Knowledge distillation is commonly used for model compression in which a small model (student model) aims to achieve approximately equivalent performance to a larger model (teacher model) by inheriting important knowledge from the larger model~\cite{hinton2015distilling}. Typically, when applying knowledge distillation in an incremental learning setting we use a \textit{teacher model} trained on historical data and a \textit{student model} trained on the incremental data block. When training the student model, a distillation term is introduced to the loss (see eq.~\eqref{eqn:general_distill_loss}), in order to retain the knowledge acquired by the teacher model. The objective function for training the incremental learning model can be formulated as: 
\begin{equation}
    \mathcal{L}_{S} = \mathcal{L}_{new}(\boldsymbol{y}_S, \boldsymbol{\tilde{y}}_S) + \lambda\mathcal{L}_{KD}(\boldsymbol{\tilde{y}}_T, \boldsymbol{\tilde{y}}_S)\,,
    \label{eqn:general_distill_loss}
\end{equation}
where $\mathcal{L}_{new}$ denotes the student model's loss function between ground truth labels $\boldsymbol{y}_S$ and the predicted values $\boldsymbol{\tilde{y}}_S$, and $\mathcal{L}_{KD}$ denotes the knowledge distillation loss between the teacher and the student models. Here, $\lambda$ denotes a scalar that controls the amount of distillation loss involved during training~\cite{hinton2015distilling}. 

In the context of recommender systems, $\mathcal{L}_{new}$ is often a Bayesian personalized ranking  (BPR)~\cite{RendleFGS2009_bpr} loss. In a recent GNN-based knowledge distillation approach called GraphSAIL~\cite{xu2020graphsail} $\mathcal{L}_{KD}$ consists of self-embedding distillation, global structure distillation, and local structure distillation. 
Though previous works have achieved promising performance and efficient training processes~\cite{xu2020graphsail, Wang2021GraphSA,gag_streaming_gnn}, they preserve previous knowledge using a single imitation weight $\lambda$ that applies to all users. However, fully preserving the historical information for all users is sub-optimal since it fails to take into account the dynamics of each user's change of preferences. In practice, we observe that user preferences are dynamic and different users can be expected to exhibit different levels of interest change between the past time blocks and the arrival of the incremental data block. For users whose interests shift significantly from previous time periods, we do not want the student model's learning on the incremental data to be overly constrained by the teacher model. Therefore, following this intuition, we aim to distinguish different distillation levels for different users, so we propose to adaptively learn an imitation weight $w_u$ for each user and apply it to the knowledge distillation objective function. We elaborate on the process for learning $w_u$ in the following subsections. 

Our personalized knowledge distillation loss is:
\begin{equation}
\mathcal{L}_{\text{KD - PIW}}^{U} = \sum_{u\in\mathcal{U}} \boldsymbol{w}_u\mathcal{L}_{KD}^{u}(\boldsymbol{\tilde{y}}_{u,T}, \boldsymbol{\tilde{y}}_{u,S}) \nonumber
\end{equation}
where $\boldsymbol{w}_u$ is the personalized imitation weight learned for each user to identify the amount of knowledge to retain from the teacher model. $\mathcal{L}_{KD}^{u}$ refers to the knowledge distillation loss for a user $u$. 

The overall incremental learning training objective is:
\begin{equation}
    \mathcal{L}_{S} = \mathcal{L}_{new}(\boldsymbol{y}_S, \boldsymbol{\tilde{y}}_S) + \lambda\left( \mathcal{L}_{\text{KD - PIW}}^{U} + \mathcal{L}_{\text{KD }}^{I}\right)
    \label{eqn: adaptive_distill_loss}
\end{equation}
where$\mathcal{L}_{\text{KD }}^{I} = \sum_{i\in\mathcal{I}}  \mathcal{L}_{KD}^{i}(\boldsymbol{\tilde{y}}_{i,T}, \boldsymbol{\tilde{y}}_{i,S})$ is the knowledge distillation loss among all the items
and $\mathcal{L}_{KD}^{i}$ refers to knowledge distillation loss for a item $i$. Since item knowledge is usually static by nature, we adopt the original knowledge distillation without imitation weight personalization. 

\subsection{Imitation Weight Generator}
Directly learning the personalized imitation weight $\boldsymbol{w}_u$ will introduce a large number of learnable parameters because of the large user number. Instead, we propose to learn the personalized imitation weights using a learnable function parameterized by a neural network. For each training block, we associate with each user a state vector $\boldsymbol{s}_u \in\mathbb{R}^M$ which represents the user interest change between consecutive time blocks. Here $M$ is the total number of clusters of items. We explain how we construct, initialize and update this state vector below. For now, assuming we have such a vector $\boldsymbol{s}_u$, we apply it as input to a weight generator network $f(\boldsymbol{s}_u)$ to learn the imitation weight of each user for distillation. In our realization of the framework, $f(\boldsymbol{s}_u)$ is specified by the following equations:
\begin{align}
    & \boldsymbol{z_{u}} = \text{relu}(\boldsymbol{W}_{1} \cdot \boldsymbol{s}_u+b_{1}),  \label{eqn: weight_generator}\\
    & w_{u} = \text{softplus}(\boldsymbol{W}_{2} \cdot \boldsymbol{z}_{u} + b_{2}),
\end{align}
with $\boldsymbol{W}_1\in\mathbb{R}^{M \times l}$, $\boldsymbol{W}_2\in\mathbb{R}^l$, $\boldsymbol{b}_1\in\mathbb{R}^l$, and $\boldsymbol{b}_2\in\mathbb{R}$ as the learnable parameters, where $l$ is size of the hidden layer.
We note that applying a more advanced network as the weight generator may further improve the performance, but this is not the focus of this work. Thus, we use a simple form where the weight generator is parameterized by a 2-layer multi-layer perceptron (MLP). In the final layer, we adopt a softplus~\cite{Zheng2015ImprovingDN} activation function to produce a strictly non-negative imitation weight. An additional normalization is applied across all the user imitation weights in each training mini-batch to better enhance stability during the training process. 

\subsection{Structure Aware Item Clustering} \label{sec:sub_state}
To derive the interest shift state vector, it is first necessary to cluster the items.
Inspired by the deep structure clustering method~\cite{attribuedCL,SDCN}, where both node attributes and higher-order structural information are fully considered during the clustering process, we adopt a similar objective function to learn the item center clusters given the underlying user-item bipartite graph.

We measure the similarity between item embeddings and the item cluster centers' embeddings. Let $q_{i,m}^{t} \in \mathbb{R}^{1}$ denote the  similarity between the item embedding from the final layers of the GNN backbone encoder $\boldsymbol{h}_{i}^{t} \in \mathbb{R}^{d}$ at time point $t$ and the item cluster center embedding (item anchor) $\boldsymbol{\mu}_m^{t} \in \mathbb{R}^{d}$. 
We measure this distance using a Student’s t-distribution to handle differently scaled clusters in a computationally convenient manner~\cite{van2008visualizing}. This can be seen as a soft clustering assignment distribution of each item. The distribution mass of item $i$ at current time block $t$ for item cluster $m$ is calculated as:
\begin{equation}
            q_{i,m}^{t} = \frac{(1+||\boldsymbol{h}_{i}^{t}-\boldsymbol{\mu}_m^{t}||^2/\nu)^{-\frac{\nu+1}{2}}}{\sum_{m' \in M}(1+||\boldsymbol{h}_{i}^{t}-\boldsymbol{\mu}_{m'}^{t}||^2/\nu)^{-\frac{\nu+1}{2}}}\,,
            \label{eqn: soft_member}
\end{equation} 
where $M$ is the total number of item clusters.


The deep structural clustering model we adopt is trained by a self-supervised learning loss as follows:
\begin{align}
    & \mathcal{L}_{soft}^{kl}=D_{KL}(P||Q)^{t}=\sum_i\sum_mp_{i,m}^{t}log\frac{p_{i,m}^{t}}{q_{i,m}^{t}}\,,\\
    &p_{i,m}^{t}=\frac{({q_{i,m}^{t}})^2/f_m^{t}}{\sum_{m' \in M}({q_{i,m'}^{t}})^2/f_{m'}^{t}}\,,
\end{align}
where $f_{m}^{t}=\sum_i q_{i,m}^{t}$, $D_{KL}$ denotes the Kullback–Leibler (KL) divergence~\cite{KLDivergence} and $p_{i}^{t} \in \mathbb{R}^{M}$ is the target distribution for item $i$ at time point $t$ which strives to push the representations closer to cluster centers. With the clusters defined, we can now derive $\mathbf{s}_u$.

\subsection{User interest shift modelling} \label{sec:sub_item_cluster}
In this section, we detail the generation and initialization of $\mathbf{s}_u$. We use $\boldsymbol{\mu}_m^{t}$ as an item cluster anchor embedding. We calculate the user distance to clusters as:
\begin{align}
      &\tilde{\boldsymbol{G}}_{u}^{t} = [\boldsymbol{\mu}_1^{t}\boldsymbol{W}_1 (\boldsymbol{h}_{u}^{t})^T,...,\boldsymbol{\mu}_M^{t}\boldsymbol{W}_M (\boldsymbol{h}_{u}^{t})^T]\,, \\
        & \boldsymbol{G}_{u,m}^{t} = \frac{e^{\tilde{\boldsymbol{G}}_{u,m}^{t}}}{\sum_{m'=1}^{M}e^{\tilde{\boldsymbol{G}}_{u,m'}^{t}}}\,,
\label{eqn: dist_G}
\end{align}
where $\boldsymbol{W}_m \in \mathbb{R}^{d \times d}$ is a cluster-specific transformation matrix. Similarly, using the node embeddings from the previous time block we can get $\boldsymbol{G}_{u}^{t-1}$. We hypothesize that user interest shift is strongly related to the change of the users' distribution of distances to item clusters. Therefore, we design the state vector with the assumption that the importance weights of users are related to their distribution changes between two time blocks. We define the state vector as: 
\begin{equation}
  \boldsymbol{s}_u = (\boldsymbol{G}_{u}^{t-1}-\boldsymbol{G}_{u}^{t}) \odot   (\boldsymbol{G}_{u}^{t-1}-\boldsymbol{G}_{u}^{t})
\end{equation}
where $\odot $ denotes the Hadamard product.

Our framework is compatible with any graph-based recommender system incremental learning architecture such as GraphSAIL~\citep{xu2020graphsail}, SGCT~\citep{Wang2021GraphSA} and LWC-KD~\cite{Wang2021GraphSA}. These are standard state-of-the-art incremental learning approaches for recommender systems. 

\subsection{The Overall Training Framework} 

Having illustrated the detailed design of the adaptive knowledge distillation loss as well as the state vector $\mathbf{s}_u$ which characterizes the user interest shift behavior, we now focus on presenting the overall training objective function. Our model is trained in a fully end-to-end fashion where the BPR loss $\mathcal{L}_{\text{BPR}}$, which is applied on the incremental block, the personalized knowledge distillation loss for users $\mathcal{L}_{\text{KD - PIW}}^{U}$, the distillation loss for items $\mathcal{L}_{\text{KD }}^{I}$, and the self-supervised loss  $\mathcal{L}_{soft}^{kl}$ for item clustering are combined jointly as follows:  
\begin{align}\label{eqn:overall_training}
    \mathcal{L} = \mathcal{L}_{\text{BPR}} +  \lambda_1 \mathcal{L}_{soft}^{kl} + \lambda_{2}\left( \mathcal{L}_{\text{KD - PIW}}^{U} + \mathcal{L}_{\text{KD }}^{I}\right)\,.
\end{align}
Here $\lambda_1$ and $\lambda_2$ are the coefficients that balance the loss contributions between the three terms. 


\begin{table}[ht]
    \centering
       \caption{Data Statistics. Avg. \% new user and Avg. \% new item refer to the average percentage of new users/items relative to all users/items in each incremental block. }
    \label{tab:data_stat}
    \resizebox{\columnwidth}{!}{
    \begin{tabular}{l|l|l|l|l|l}
    \toprule
         \backslashbox{Stat}{Dataset} & Gowalla & Yelp & Taobao2014 & Taobao2015 & Netflix \\
         \midrule
         \# edges &281412&942395&749438&1332602&12402763\\
         \# users &5992&40863&8844&92605&63691\\
         Avg. user degrees &46.96&23.06&84.74&14.39&194.73\\
         \# items &5639&25338&39103&9842&10271\\
         Avg. item degrees & 49.90 &37.19&19.17&135.40&1207.56\\
         Avg. \% new user &2.67&3.94&1.67&2.67&4.36\\
         Avg. \% new item &0.67&1.72&2.60&0.22&0.72\\
         \# Time span (months) &19&6&1&5&6\\
    \bottomrule     
    \end{tabular}}
    \end{table}
    
\begin{table}[h]
    \caption{Performance comparison (Recall@20) of all baselines and three recent knowledge distillation algorithms with our proposed personalized adaptive weights design. The improvement ratio is with respect to fine-tune performance. }
    \label{tab:main_table}
  \resizebox{0.95\columnwidth}{!}{\begin{tabular}{c|c|cccc|c}
         \toprule
          Dataset & Methods &Inc 1 &Inc 2 &Inc 3 &Avg. &Imp \% \\
          \midrule
        \multirow{10}{*}{Gowalla} & Fine Tune &0.1412&0.1637&0.2065&0.1705&0.00 \\
        \cline{2-7}
        & LSP\_s  &0.1512&0.1741&0.2097&0.1783&4.57 \\
        \cline{2-7}
        & Uniform  &0.1480&0.1653&0.2051&0.1728&1.34 \\
        \cline{2-7}
        & Inv\_degree &0.1483&0.1680&0.2001&0.1738&1.93 \\
        \cline{2-7}
        & GraphSAIL  &0.1529&0.1823&0.2195&0.1849&8.44 \\
        & GraphSAIL-PIW &0.1547&0.1825&0.2253&0.1875&9.97 \\
        \cline{2-7}
        & SGCT &0.1588&0.1815&0.2207&0.1870&9.68 \\
        &SGCT-PIW &0.1599&0.1892&0.2321&0.1937&13.6 \\
        \cline{2-7}
        & LWC-KD&0.1639&0.1921&0.2368&0.1977&15.9 \\
        &LWC-KD-PIW &0.1698&0.1978&0.2425&\textbf{0.2033}&\textbf{19.3}\\
        \midrule
        \multirow{10}{*}{Yelp} & Fine Tune &0.0705&0.0638&0.0640&0.0661&0.00\\
        \cline{2-7}
        & LSP\_s &0.0722&0.0661&0.0644&0.0676&2.27\\
        \cline{2-7}
        & Uniform &0.0718&0.0635&0.0610&0.0654&-1.05\\
        \cline{2-7}
        &Inv\_degree &0.0727 & 0.0699 & 0.0605 & 0.0677 & 2.42\\
        \cline{2-7}
        & GraphSAIL &0.0674&0.0617&0.0625&0.0639&-3.33\\
        &GraphSAIL-PIW  &0.0718&0.0638&0.0615&0.0657&-0.66\\
        \cline{2-7}
        & SGCT &0.0740&0.0656&0.0608&0.0668&1.06\\
        &SGCT-PIW&0.0735&0.0655&0.0632&0.0674&1.92\\
        \cline{2-7}
        &LWC-KD&0.0739&0.0661&0.0637&0.0679&2.72\\
        &LWC-KD-PIW &0.0760&0.0690&0.0651&\textbf{0.0700}&\textbf{5.95}\\
        \midrule
        \multirow{10}{*}{Taobao14} & Fine Tune &0.0208&0.0112&0.0138&0.0153&0.00\\
        \cline{2-7}
        & LSP\_s &0.0213&0.0106&0.0138&0.0152&-0.65\\
        \cline{2-7}
        & Uniform &0.0195&0.0127&0.0148&0.0157&2.61\\
        \cline{2-7}
        &Inv\_degree & 0.0228 &0.0140 & 0.0159 & 0.0175 & 14.63\\
        \cline{2-7}
        & GraphSAIL &0.0222&0.0105&0.0139&0.0155&1.31\\
        &GraphSAIL-PIW &0.0206&0.0103&0.0129&0.0146&-4.58\\
        \cline{2-7}
        & SGCT &0.0240&0.0092&0.0148&0.0160&1.74\\
        &SGCT-PIW &0.0227&0.0104&0.0142&0.0158&3.05\\
        \cline{2-7}
        & LWC-KD&0.0254&0.0119&0.0156&0.0176&15.3\\
        &LWC-KD-PIW &0.0256&0.0118&0.0161&\textbf{0.0178}&\textbf{16.3}\\
        \midrule
        \multirow{10}{*}{Taobao15} & Fine Tune &0.0933&0.0952&0.0965&0.0950&0.00\\
        \cline{2-7}
        & LSP\_s &0.0993&0.0952&0.0957&0.0968&1.86\\
        \cline{2-7}
        & Uniform &0.0988&0.0954&0.1004&0.0982&3.37\\
        \cline{2-7}
        & Inv\_degree &0.0991 &0.0977 & 0.1000 & 0.0989 & 4.16\\
        \cline{2-7}
        & GraphSAIL &0.0959&0.0959&0.0972&0.0963&1.39\\
        & GraphSAIL-PIW &0.1024&0.0983&0.1018&0.1008&6.14\\
        \cline{2-7}
        & SGCT &0.1030&0.0983&0.0984&0.0999&5.16\\
        &SGCT-PIW &0.1040&0.0999&0.1027&0.1022&7.58\\
        \cline{2-7}
        & LWC-KD&0.1039&0.1022&0.1029&0.1030&8.42\\
        & LWC-KD-PIW &0.1044&0.1045&0.1052&\textbf{0.1047}&\textbf{10.2}\\
        \midrule

        \multirow{9}{*}{Netflix} & Fine Tune  &0.1092&0.1041&0.0977&0.1036&0.00\\
        \cline{2-7}
        & LSP\_s &0.1173&0.1136&0.1076&0.1128&8.88\\
        \cline{2-7}
        & Uniform  &0.1018&0.1055&0.0800&0.0957&-7.63\\
        \cline{2-7}
        & Inv\_degree &0.1000&0.1050&0.0820&0.0957&-7.63\\
        \cline{2-7}
        & GraphSAIL &0.1163&0.1023&0.0968 & 0.1051&1.45\\
        &GraphSAIL-PIW &0.1142&0.1028&0.0986&0.1052&1.54\\
        \cline{2-7}
        & SGCT &0.1166&0.1161&0.1077&0.1135&9.56\\
        &SGCT-PIW &0.1185&0.1144&0.1098&0.1142&10.23\\
        \cline{2-7}
        & LWC-KD&0.1185&0.1170&0.1071&\textbf{0.1142}&\textbf{10.23}\\
        &LWC-KD-PIW &0.1185&0.1146&0.1087&0.1139&9.97\\
        \bottomrule
     \end{tabular}}
     \vspace{-1em}
\end{table}
\section{EXPERIMENTS}\label{experiment}
\subsubsection{Datasets}
We use a diverse set of datasets consisting of real-world user-item interactions. As shown in Table \ref{tab:data_stat}, the datasets vary in the number of edges and number of user and item nodes by up to two orders of magnitude, demonstrating our approach's scalability. 
The 5 mainstream, publicly available datasets we use are: Gowalla, Yelp, Taobao14, Taobao15 and Netflix. 

\subsubsection{Base Model}

We use MGCCF~\cite{sun2020_mgcf} as our base model in the incremental learning methods. It is a state-of-art backbone model in the incremental recommendation framework which not only incorporates multiple graphs in the embedding learning process, but also considers the intrinsic difference between user nodes and item nodes when performing graph convolution on the bipartite graph. 

\subsubsection{Baselines} To demonstrate that our model's strength, we compare our algorithm with multiple baselines.

\noindent\textbf{Fine Tune}: Fine Tune uses solely the new data of each time block to fine-tune the model that was trained using the previous time blocks.

\noindent\textbf{LSP\_s}~\cite{yang2020distilling}: LSP is a recent state-art-of approach which applies knowledge distillation on Graph Convolution Network (GCN) models. It preserves local structure from the teacher to student by minimizing the distances between distributions representing local topological semantics.

\noindent\textbf{Uniform}: This is a naive reservoir replay method. A subset of old data is sampled and added to the new data.

\noindent\textbf{Inv\_degree}~\cite{Ahrabian2021StructureAE}: Inv\_degree is a state-of-art reservoir replay method. The reservoir is based on graph structure. The approach selects a fixed-size subset of user-item pairs from historical data; each interaction's selection probability is proportional to the inverse degree of its user.  

\noindent\textbf{SOTA Graph Rec. Sys. Incremental Learning methods}: GraphSail~\cite{xu2020graphsail}, SGCT~\cite{Wang2021GraphSA} and LWC-KD~\cite{Wang2021GraphSA} are state-of-the-art models which we improve upon by integrating our approach. To demonstrate the strength of our method we compare with the base models.
\begin{figure}[ht]
    \includegraphics[width=0.49\textwidth]{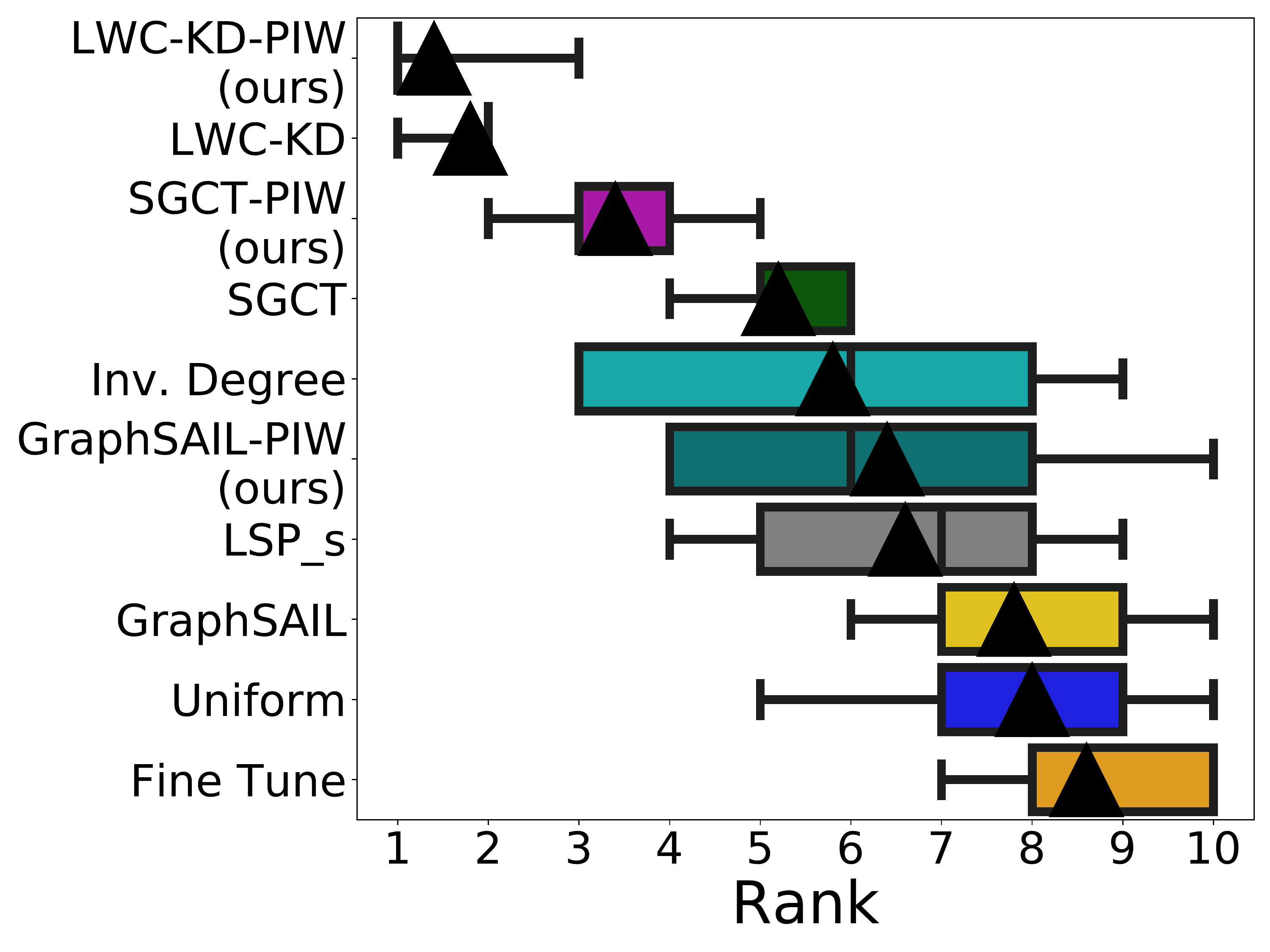}
    \caption{Boxplot of ranks of the algorithms across the 5 datasets. The medians and means of the ranks are shown by the vertical lines and the black triangles respectively; whiskers extend to the minimum and maximum ranks.} 
    \label{fig:box_plot}
\end{figure}
\begin{figure*}[htp]
    \vspace{-1em}

    \centering
    \includegraphics[width=0.9\textwidth]{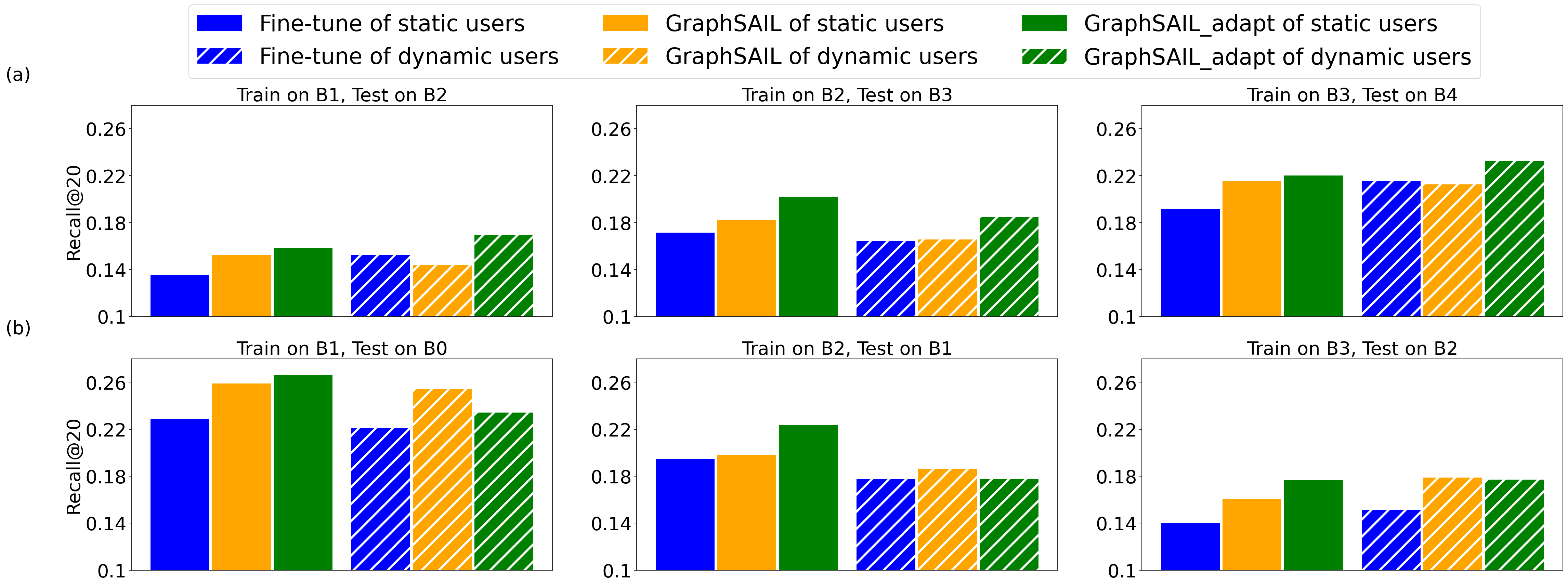}

        \caption{\textbf{Case study}: \textbf{(a)} Model performance for two groups of users: static users (20\% of users whose interests shift the least) and dynamic users (20\% of users whose interests shift most). For both groups, GraphSAIL\_adapt outperforms fine-tune and GraphSAIL w/o adaptive weights. To explore how adaptive weights effectively control the amount of knowledge distilled from the teacher model for each user, we evaluate models by training on block t and testing on block t-1  in \textbf{(b)}. GraphSAIL\_adapt outperforms GraphSAIL for the static user group, while GraphSAIL performs better for the dynamic user group. This indicates that GraphSAIL\_adapt preserves more information for users with persistent preferences and forgets more for users with the most dynamic preferences. Therefore, different levels of distillation help to improve the performance of the student model.
        }
        \label{fig:case_study_gowalla}
\end{figure*}
\subsection{Results and Discussion}
Table \ref{tab:main_table} reports the performance of baselines and three distillation algorithms with/without adaptive weights along with standard reservoir replay methods. Please note that all the results reported are an average across three trails with different random seeds. The results across all datasets of Table~\ref{tab:main_table} are summarized in Figure \ref{fig:box_plot}. 
We note that our adaptive weight framework improves method performance in all cases, since the baseline methods for LWC-KD, SGCT and GraphSAIL improve their average and median rank when the adaptive weights are incorporated. 
Our methods are routinely in the top three algorithms for all datasets and often achieve the best performance (see last column of Table~\ref{tab:main_table}).
Besides the relative rank, in terms of absolute performance gain, the adaptive weights provide double digit percent performance increase over fine tuning across a variety of datasets. 
In particular, the strongest performance of adaptive weights is observed in traditional incremental learning datasets such as Gowalla. 
In datasets such as Netflix most users have a very high number of interactions (more than 100). As a result, the base portion of the dataset provides a good reflection of each user's set of interests. Users are therefore less likely to exhibit drastic changes of interest.

\textbf{Comparison with full-batch training} With Gowalla and Taobao2014, we have trained using all previous blocks and block t, and tested on block t+1 for each incremental block t (i.e. full batch). The average recall@20 is 0.1963 for Gowalla and 0.0191 for Taobao14. Though obtaining better performance in Taobao2014, the full batch method takes three times more time to train compared to LWC-KD-PIW. Therefore, in a live deployment setting for a client-facing recommender system, our system would be able to provide daily updates, whereas full retraining would quickly become computationally infeasible as new data accumulated.

\subsection{Case study}

We have conducted a case study in order to more closely examine how three models behave for two distinct groups of users. The three models we study are fine-tune, GraphSAIL without adaptive weights and GraphSAIL with adaptive weights. We identify two types of users: (i) 
{\em static} users who exhibit minimal interest shift; and {\em dynamic} users who exhibit dramatic interest shift. Then we test the models on the historical data (i.e., data from previous time block). This evaluation on old data provides us with insight into how much historical information the models preserve for each group of users. We also check how well each group of users performs by testing on the next time block. Therefore, we can assess how preserving different amounts of information for each user group affects the model performance on the task of interest. 

 We observe that GraphSAIL with adaptive weights performs best for both user groups (Figure \ref{fig:case_study_gowalla} (a)). From evaluation on historical data (Figure \ref{fig:case_study_gowalla} (b)), we see that GraphSAIL is less affected than fine-tune, indicating that it counters the forgetting problem. GraphSAIL with adaptive weights outperforms GraphSAIL without adaptive weights for static users, while GraphSAIL without adaptive weights performs better with dynamic users. This implies that GraphSAIL with adaptive weights preserves more information for static users and less for dynamic users. Therefore, we conclude that adaptively distilling knowledge can help to improve modelling of future user preferences.

\section{Conclusion}
In this paper, we have proposed a novel method for incremental learning in graph-based recommender systems. Our approach hinges on learning adaptive personalization weights to tune the amount of knowledge distilled for each user's preferences during incremental learning.
Our proposed method is evaluated on multiple datasets with three different incremental learning backbones and it consistently outperforms standard non-adaptive techniques.
Our case study further supports our claim that the use of adaptive weights allows the model to distill more information for users with constant interests and to retain less information for users that are expressing rapid change in interests. This allows the model to adapt more quickly to changes of preferences for users with evolving interests.

\bibliography{main}

\begin{thebibliography}{35}
\providecommand{\natexlab}[1]{#1}

\bibitem[{Ahrabian et~al.(2021)Ahrabian, Xu, Zhang, Wu, Wang, and
  Coates}]{Ahrabian2021StructureAE}
Ahrabian, K.; Xu, Y.; Zhang, Y.; Wu, J.; Wang, Y.; and Coates, M. 2021.
\newblock Structure Aware Experience Replay for Incremental Learning in
  Graph-based Recommender Systems.
\newblock In \emph{Proc. ACM Int. Conf. Info. \& Knowledge Management},
  2832–2836.

\bibitem[{Bo et~al.(2020)Bo, Wang, Shi, Zhu, Lu, and Cui}]{SDCN}
Bo, D.; Wang, X.; Shi, C.; Zhu, M.; Lu, E.; and Cui, P. 2020.
\newblock Structural Deep Clustering Network.
\newblock In \emph{Proc. World Wide Web Conf. (WWW)}, 1400--1410.

\bibitem[{Castro et~al.(2018)Castro, Mar{\'{\i}}n{-}Jim{\'{e}}nez, Guil,
  Schmid, and Alahari}]{Castro2018E2E}
Castro, F.~M.; Mar{\'{\i}}n{-}Jim{\'{e}}nez, M.~J.; Guil, N.; Schmid, C.; and
  Alahari, K. 2018.
\newblock End-to-End Incremental Learning.
\newblock In \emph{Proc. European Conf. Computer Vision (ECCV)}, 241--257.

\bibitem[{Cheng et~al.(2016)Cheng, Koc, Harmsen, Shaked, Chandra
  et~al.}]{cheng2016wide}
Cheng, H.; Koc, L.; Harmsen, J.; Shaked, T.; Chandra, T.; et~al. 2016.
\newblock Wide {\&} Deep Learning for Recommender Systems.
\newblock In \emph{Proc. ACM Recommender Syst. Conf. - Workshop on Deep
  Learning for Recommender Syst.}, 7--10.

\bibitem[{Chiang et~al.(2019)Chiang, Liu, Si, Li, Bengio, and
  Hsieh}]{chiang2019cluster}
Chiang, W.-L.; Liu, X.; Si, S.; Li, Y.; Bengio, S.; and Hsieh, C.-J. 2019.
\newblock Cluster-gcn: An efficient algorithm for training deep and large graph
  convolutional networks.
\newblock In \emph{Proc. ACM SIGKDD Int. Conf. Knowledge Discovery \& Data
  Mining}, 257--266.

\bibitem[{Covington, Adams, and Sargin(2016)}]{covington2016deep}
Covington, P.; Adams, J.; and Sargin, E. 2016.
\newblock Deep neural networks for youtube recommendations.
\newblock In \emph{Proc. ACM Conf. recommender systems}, 191--198.

\bibitem[{Fan et~al.(2021)Fan, Liu, Zhang, Xiong, Zheng, and
  Yu}]{fan2021continuous}
Fan, Z.; Liu, Z.; Zhang, J.; Xiong, Y.; Zheng, L.; and Yu, P.~S. 2021.
\newblock Continuous-time sequential recommendation with temporal graph
  collaborative transformer.
\newblock In \emph{CIKM}.

\bibitem[{Guo et~al.(2017)Guo, Tang, Ye, Li, and He}]{deepfm}
Guo, H.; Tang, R.; Ye, Y.; Li, Z.; and He, X. 2017.
\newblock Deepfm: a factorization-machine based neural network for ctr
  prediction.
\newblock In \emph{Proc. Int. Joint. Conf. Artificial Intelligence (IJCAI)},
  1725--1731.

\bibitem[{He et~al.(2020)He, Deng, Wang, Li, Zhang, and Wang}]{he2020lightgcn}
He, X.; Deng, K.; Wang, X.; Li, Y.; Zhang, Y.; and Wang, M. 2020.
\newblock LightGCN: Simplifying and Powering Graph Convolution Network for
  Recommendation.
\newblock \emph{Proc. ACM Int. Conf. Research and Development in Info.
  Retrieval}.

\bibitem[{Hidasi and Karatzoglou(2018)}]{hidasi2018recurrent}
Hidasi, B.; and Karatzoglou, A. 2018.
\newblock Recurrent neural networks with top-k gains for session-based
  recommendations.
\newblock In \emph{CIKM}.

\bibitem[{Hinton, Vinyals, and Dean(2015)}]{hinton2015distilling}
Hinton, G.; Vinyals, O.; and Dean, J. 2015.
\newblock Distilling the Knowledge in a Neural Network.
\newblock \emph{CoRR}, arXiv.

\bibitem[{Kang and McAuley(2018)}]{kang2018self}
Kang, W.-C.; and McAuley, J. 2018.
\newblock Self-attentive sequential recommendation.
\newblock In \emph{ICDM}.

\bibitem[{Kirkpatrick et~al.(2017)Kirkpatrick, Pascanu, Rabinowitz, Veness,
  Desjardins, Rusu, Milan, Quan, Ramalho, Grabska-Barwinska
  et~al.}]{kirkpatrick2017overcoming}
Kirkpatrick, J.; Pascanu, R.; Rabinowitz, N.; Veness, J.; Desjardins, G.; Rusu,
  A.~A.; Milan, K.; Quan, J.; Ramalho, T.; Grabska-Barwinska, A.; et~al. 2017.
\newblock Overcoming catastrophic forgetting in neural networks.
\newblock \emph{Proc. national academy of sciences}, 114(13): 3521--3526.

\bibitem[{Kullback and Leibler(1951)}]{KLDivergence}
Kullback, S.; and Leibler, R.~A. 1951.
\newblock {On Information and Sufficiency}.
\newblock \emph{The Annals of Mathematical Statistics}, 22: 79 -- 86.

\bibitem[{Mallya and Lazebnik(2018)}]{mallya2018packnet_IL}
Mallya, A.; and Lazebnik, S. 2018.
\newblock Packnet: Adding multiple tasks to a single network by iterative
  pruning.
\newblock In \emph{Conf. Computer Vision and Pattern Recognition (CVPR)},
  7765--7773.

\bibitem[{Prabhu, Torr, and Dokania(2020)}]{prabhu2020gdumb}
Prabhu, A.; Torr, P.~H.; and Dokania, P.~K. 2020.
\newblock GDumb: A simple approach that questions our progress in continual
  learning.
\newblock In \emph{European Conf. Computer Vision (ECCV)}, 524--540.

\bibitem[{Qiu et~al.(2020)Qiu, Yin, Huang, and Tong}]{gag_streaming_gnn}
Qiu, R.; Yin, H.; Huang, Z.; and Tong, C. 2020.
\newblock GAG: Global Attributed Graph Neural Network for Streaming
  Session-based Recommendation.
\newblock In \emph{Int. ACM SIGIR Conf. Research and Development in Info.
  Retrieval}, 669--678.

\bibitem[{Rebuffi et~al.(2017)Rebuffi, Kolesnikov, Sperl, and
  Lampert}]{rebuffi2017icarl}
Rebuffi, S.-A.; Kolesnikov, A.; Sperl, G.; and Lampert, C.~H. 2017.
\newblock iCarl: Incremental classifier and representation learning.
\newblock In \emph{Proc. IEEE Conf. Computer Vision and Pattern Recognition},
  2001--2010.

\bibitem[{Rendle et~al.(2009)Rendle, Freudenthaler, Gantner, and
  Schmidt{-}Thieme}]{RendleFGS2009_bpr}
Rendle, S.; Freudenthaler, C.; Gantner, Z.; and Schmidt{-}Thieme, L. 2009.
\newblock {BPR:} Bayesian Personalized Ranking from Implicit Feedback.
\newblock In \emph{Proc. Conf. Uncertainty in Artificial Intell. (UAI)},
  452--461.

\bibitem[{Shmelkov, Schmid, and Alahari(2017)}]{Shmelkov_2017_ICCV}
Shmelkov, K.; Schmid, C.; and Alahari, K. 2017.
\newblock Incremental Learning of Object Detectors Without Catastrophic
  Forgetting.
\newblock In \emph{Proc. Int. Conf. Computer Vision (ICCV)}, 3400--3409.

\bibitem[{Sun et~al.(2019)Sun, Zhang, Ma, Coates, Guo, Tang, and
  He}]{sun2020_mgcf}
Sun, J.; Zhang, Y.; Ma, C.; Coates, M.; Guo, H.; Tang, R.; and He, X. 2019.
\newblock Multi-Graph Convolution Collaborative Filtering.
\newblock In \emph{Proc. IEEE Int. Conf. Data Mining (ICDM)}, 1306--1311.

\bibitem[{van~den Berg, Kipf, and Welling(2017)}]{gcmc_vdberg2018}
van~den Berg, R.; Kipf, T.~N.; and Welling, M. 2017.
\newblock Graph Convolutional Matrix Completion.
\newblock volume abs/1706.02263.

\bibitem[{Van~der Maaten and Hinton(2008)}]{van2008visualizing}
Van~der Maaten, L.; and Hinton, G. 2008.
\newblock Visualizing data using t-SNE.
\newblock \emph{J. machine learning research}, 9: 2579--2605.

\bibitem[{Wang et~al.(2019{\natexlab{a}})Wang, Pan, Hu, Long, Jiang, and
  Zhang}]{attribuedCL}
Wang, C.; Pan, S.; Hu, R.; Long, G.; Jiang, J.; and Zhang, C.
  2019{\natexlab{a}}.
\newblock Attributed Graph Clustering: {A} Deep Attentional Embedding Approach.
\newblock In \emph{Proc. Int. Joint. Conf. Artificial Intelligence (IJCAI)},
  3670--3676.

\bibitem[{Wang et~al.(2021)Wang, Hu, Wang, He, Sheng, Orgun, Cao, Ricci, and
  Yu}]{wang2021graph}
Wang, S.; Hu, L.; Wang, Y.; He, X.; Sheng, Q.~Z.; Orgun, M.~A.; Cao, L.; Ricci,
  F.; and Yu, P.~S. 2021.
\newblock Graph learning based recommender systems: A review.
\newblock In \emph{Proc. Int. Joint. Conf. Artificial Intelligence (IJCAI)},
  4644--4652.

\bibitem[{Wang et~al.(2019{\natexlab{b}})Wang, He, Wang, Feng, and
  Chua}]{NGCF_wang19}
Wang, X.; He, X.; Wang, M.; Feng, F.; and Chua, T.-S. 2019{\natexlab{b}}.
\newblock Neural Graph Collaborative Filtering.
\newblock In \emph{Proc. ACM Int. Conf. Research and Development in Info.
  Retrieval}, 165--174.

\bibitem[{Wang, Zhang, and Coates(2021)}]{Wang2021GraphSA}
Wang, Y.; Zhang, Y.; and Coates, M. 2021.
\newblock Graph Structure Aware Contrastive Knowledge Distillation for
  Incremental Learning in Recommender Systems.
\newblock In \emph{Proc. ACM Int. Conf. Info. \& Knowledge Management},
  3518--3522.

\bibitem[{Xu and Zhu(2018)}]{xu2018reinforced_IL}
Xu, J.; and Zhu, Z. 2018.
\newblock Reinforced continual learning.
\newblock In \emph{Proc. Adv. Neural Info. Syst. (NeurIPS)}, 907--916.

\bibitem[{Xu et~al.(2020)Xu, Zhang, Guo, Guo, Tang, and
  Coates}]{xu2020graphsail}
Xu, Y.; Zhang, Y.; Guo, W.; Guo, H.; Tang, R.; and Coates, M. 2020.
\newblock GraphSAIL: Graph Structure Aware Incremental Learning for Recommender
  Systems.
\newblock In \emph{Proc. ACM Int. Conf. Info. \& Knowledge Management},
  2861--2868.

\bibitem[{Yang et~al.(2020)Yang, Qiu, Song, Tao, and Wang}]{yang2020distilling}
Yang, Y.; Qiu, J.; Song, M.; Tao, D.; and Wang, X. 2020.
\newblock Distilling Knowledge from Graph Convolutional Networks.
\newblock In \emph{Proc. IEEE Conf. Computer Vision and Pattern Recognition},
  7074--7083.

\bibitem[{Yang et~al.(2019)Yang, Zhou, Zhan, Xiong, and
  Jiang}]{yang2019adaptive}
Yang, Y.; Zhou, D.-W.; Zhan, D.-C.; Xiong, H.; and Jiang, Y. 2019.
\newblock Adaptive Deep Models for Incremental Learning: Considering Capacity
  Scalability and Sustainability.
\newblock In \emph{Proc. {ACM} Conf. Knowledge Discovery \& Data Mining},
  74--82.

\bibitem[{Ying et~al.(2018)Ying, He, Chen, Eksombatchai, Hamilton, and
  Leskovec}]{ying2018}
Ying, R.; He, R.; Chen, K.; Eksombatchai, P.; Hamilton, W.~L.; and Leskovec, J.
  2018.
\newblock Graph Convolutional Neural Networks for Web-Scale Recommender
  Systems.
\newblock In \emph{Proc. {ACM} Conf. Knowledge Discovery \& Data Mining},
  974--983.

\bibitem[{Zeng et~al.(2020)Zeng, Zhou, Srivastava, Kannan, and
  Prasanna}]{graphsaint-iclr20}
Zeng, H.; Zhou, H.; Srivastava, A.; Kannan, R.; and Prasanna, V. 2020.
\newblock {GraphSAINT}: Graph Sampling Based Inductive Learning Method.
\newblock In \emph{Proc. Int. Conf. Learning Representations (ICLR)}.

\bibitem[{Zheng et~al.(2015)Zheng, Yang, Liu, Liang, and
  Li}]{Zheng2015ImprovingDN}
Zheng, H.; Yang, Z.; Liu, W.; Liang, J.; and Li, Y. 2015.
\newblock Improving deep neural networks using softplus units.
\newblock \emph{2015 International Joint Conference on Neural Networks
  (IJCNN)}, 1--4.

\bibitem[{Zou et~al.(2019)Zou, Hu, Wang, Jiang, Sun, and Gu}]{zou2019layer}
Zou, D.; Hu, Z.; Wang, Y.; Jiang, S.; Sun, Y.; and Gu, Q. 2019.
\newblock Layer-Dependent Importance Sampling for Training Deep and Large Graph
  Convolutional Networks.
\newblock In \emph{Proc. Adv. Neural Info. Syst. (NeurIPS)}, 11247--11256.

\end{thebibliography}


\begin{thebibliography}{5}
\providecommand{\natexlab}[1]{#1}

\bibitem[{Ahrabian et~al.(2021)Ahrabian, Xu, Zhang, Wu, Wang, and
  Coates}]{Ahrabian2021StructureAE}
Ahrabian, K.; Xu, Y.; Zhang, Y.; Wu, J.; Wang, Y.; and Coates, M. 2021.
\newblock Structure Aware Experience Replay for Incremental Learning in
  Graph-based Recommender Systems.
\newblock In \emph{Proc. ACM Int. Conf. Info. \& Knowledge Management},
  2832–2836.

\bibitem[{Kingma and Ba(2015)}]{kingma2015adam}
Kingma, D.~P.; and Ba, J. 2015.
\newblock Adam: A Method for Stochastic Optimization.
\newblock \emph{CoRR}, abs/1412.6980.

\bibitem[{Sun et~al.(2019)Sun, Zhang, Ma, Coates, Guo, Tang, and
  He}]{sun2020_mgcf}
Sun, J.; Zhang, Y.; Ma, C.; Coates, M.; Guo, H.; Tang, R.; and He, X. 2019.
\newblock Multi-Graph Convolution Collaborative Filtering.
\newblock In \emph{Proc. IEEE Int. Conf. Data Mining (ICDM)}, 1306--1311.

\bibitem[{Wang, Zhang, and Coates(2021)}]{Wang2021GraphSA}
Wang, Y.; Zhang, Y.; and Coates, M. 2021.
\newblock Graph Structure Aware Contrastive Knowledge Distillation for
  Incremental Learning in Recommender Systems.
\newblock In \emph{Proc. ACM Int. Conf. Info. \& Knowledge Management},
  3518--3522.

\bibitem[{Xu et~al.(2020)Xu, Zhang, Guo, Guo, Tang, and
  Coates}]{xu2020graphsail}
Xu, Y.; Zhang, Y.; Guo, W.; Guo, H.; Tang, R.; and Coates, M. 2020.
\newblock GraphSAIL: Graph Structure Aware Incremental Learning for Recommender
  Systems.
\newblock In \emph{Proc. ACM Int. Conf. Info. \& Knowledge Management},
  2861--2868.

\end{thebibliography}

\end{document}


\maketitle

\section{Additional Motivation Figure}

\begin{figure}[htb]
    \centering
    \includegraphics[width=0.9\columnwidth]{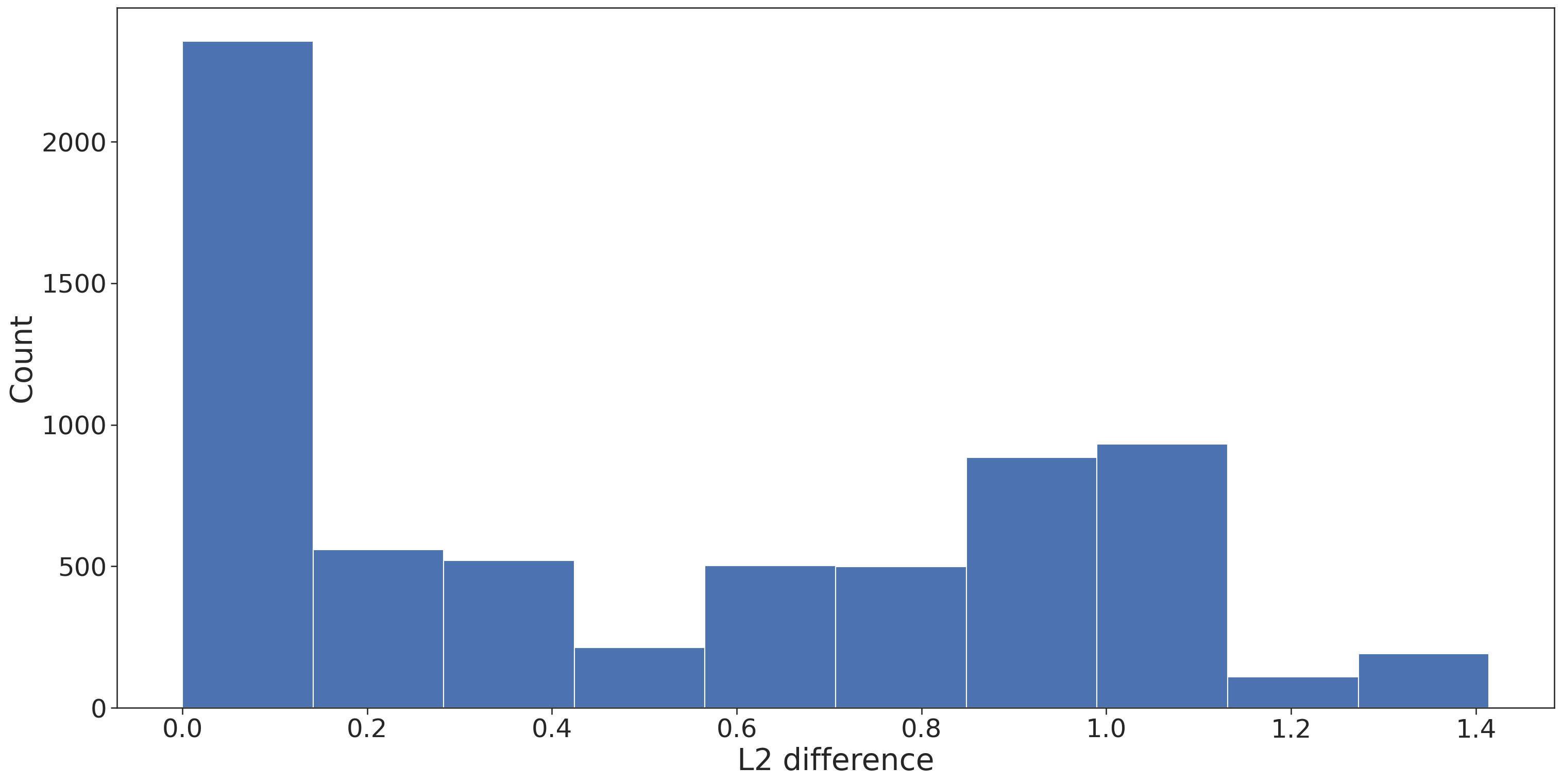}
    \caption{Illustration user interest shift between two successive blocks for Taobao 2014. }
    \label{fig:demo}
    \vspace{-1em}
\end{figure}

We believe preserving the same amount historical information for all users might be sub-optimal since it fails to take into account the dynamics of each user's potential change of preferences. For the users whose interests shift substantially, retaining too much of the old knowledge from the past via the knowledge distillation process, might constrain the model from quickly adapting to the users’ latest interests. In contrast, for users who have more static preferences, model performance can benefit greatly from preserving as much of the user's long-term preference as possible. We illustrate this with an example in Figure \ref{fig:demo}. We extract each user's interest distribution at each time block based on the interacted item's category. We define each user's probability to interact with items from category $c$ at time block $t$ as $\frac{e^{N_c^t}}{\sum_{c'=1}^Ce^{N_{c'}^t}}$, where $N_c^t$ refers to the number of items the user interacted with from category c at time block $t$ and $C$ denotes the total number of item categories. The categories are pre-defined in the dataset (tags of item). The histogram depicts the magnitude of L2 difference in user's interest distribution between the base block of the Taobao2014 dataset and the first incremental block. While most users maintain interactions with items from similar categories across time blocks, demonstrating low distribution change (left side of the plot), there do exist a portion of users with large interest distribution change between the consecutive time blocks (center \& right side of the plot). Thus, it is vital to take into account the dynamics of each user's potential change of preferences while performing the knowledge distillation-based incremental learning. 

\section{Dataset Details}

We use the experimental setup and evaluation of the baselines~\cite{Ahrabian2021StructureAE, xu2020graphsail, Wang2021GraphSA}.
\begin{itemize}
    \item \textbf{Gowalla}\footnote{https://snap.stanford.edu/data/loc-gowalla.html}: This is a real-world dataset collected from users sharing their location by checking-in on a location-based social networking website. We filtered out user and item nodes with fewer than 20 interactions.

\item \textbf{Yelp}\footnote{https://www.yelp.com/dataset}: This is a dataset of Yelp's business reviews and user data. Due to the large volume of data, we only kept the five most recent years of data and filtered out user and item nodes with fewer than 10 interactions.

\item \textbf{Taobao2014}\footnote{https://tianchi.aliyun.com/dataset/dataDetail?dataId=46}: Real user behavior data from 2014 provided by Alibaba Group. We filtered out user and item nodes with fewer than 10 interactions.  

\item \textbf{Taobao2015}\footnote{https://tianchi.aliyun.com/dataset/dataDetail?dataId=53}: This real-world dataset accumulated on Tmall/ Taobao and the app Alipay in 2015. We filtered out user and item nodes with fewer than 10 interactions.

\item \textbf{Netflix-Prize}\footnote{https://academictorrents.com/details/\\9b13183dc4d60676b773c9e2cd6de5e5542cee9a}: This is the official user's movie rating dataset used in the Netflix Prize competition. We only kept interactions with rating of 5. We filtered out user and item nodes with fewer than 100 interactions. 
\end{itemize}

\begin{figure}[htb]
    \centering
    \includegraphics[width=0.9\columnwidth]{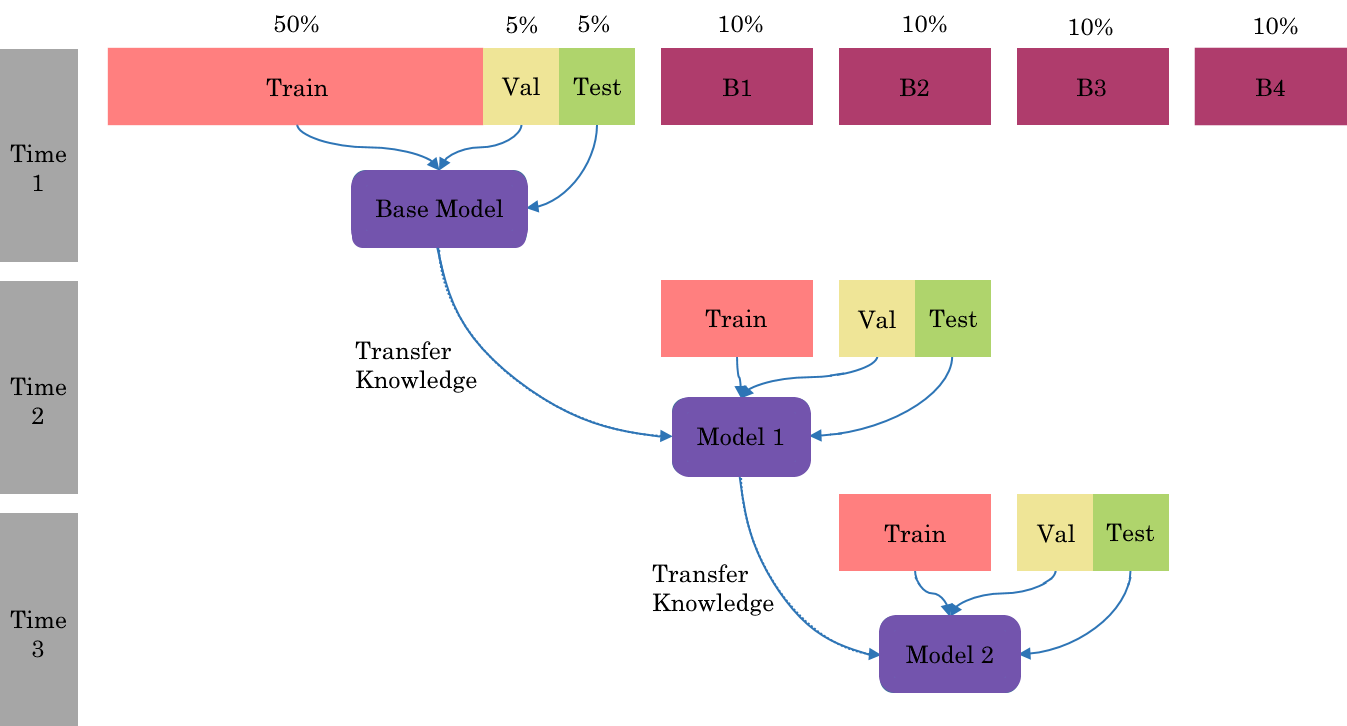}
    \caption{Data Split of experimental setting.}
    \label{fig:data_splitting}
\end{figure}

For all datasets, we split the data in chronological order into a base block and incremental blocks. The base block contains 60\% of the data and is randomly split into training, validation and testing sets. The remaining 40\% of the data is evenly separated into four incremental blocks. During training, when block $t$ is used as the training set, the first half of block $t+1$ is used as the validation set and the second half is used as the testing set (Figure \ref{fig:data_splitting}).

\section{Hyperparameter Settings} \label{app:hyperparameter_settings}

The training loop is implemented in TensorFlow using Adam optimizer~\cite{kingma2015adam}. The learning rate is fixed at 0.0005. We train the model for a minimum of 10 epochs, with early stopping and a patience parameter of 2.

The backbone graph neural network is the MGCCF~\cite{sun2020_mgcf} trained using the hyperparameters shown in Table~\ref{table:hyperparameter_settings}.

\begin{table}[!ht]
    \centering
        \caption{Hyperparameters of our model on all benchmarks.}
    \begin{tabular}{lc}
        \toprule
        Hyperparameter & Value  \\ 
        \midrule
        Min Epochs Base Block	& 10	\\
        Min Epochs Incremental	& 3	\\ 
        Max Epochs Base Block	& N/A	\\ 
        Max Epochs Incremental	& 10	\\ 
        Early Stopping Patience	& 2	\\ 
        Batch size & 64	\\
        Optimizer & Adam \\
        Learning rate (max) & 5e-4  \\
        Dropout & 0.2 \\
        Losses & L2, BPR \\
        GNN Num Layers ($R$) & 2 \\
        Embedding dimensionality & 128 \\
        Augmentations & NONE \\
        \bottomrule
    \end{tabular}
    \vspace{1em}
    \label{table:hyperparameter_settings}
\end{table}

\section{Computation Requirements \& Empirical Time Complexity}\label{sec:compute}

For the biggest dataset in our experiment, Netflix, on our dedicated server with 72 Intel Xeon (R) Gold 6140 2.30GHz CPUs and an NVIDIA Tesla V100 GPU, it takes approximately 40 hours to train the base block which represents 60\% of the dataset. Each incremental block comprises 10\% of the dataset and trains at about 7 hours. If we were to run full batch re-training at each time step, training for all 4 blocks would require approximately $40+47+54+61=202$ hours. On the other hand, using the incremental learning approach, training lasts $40+7+7+7=61$ hours (a 70\% reduction!). This result shows that our incremental learning design has significant time and energy savings potential. 

On the model accuracy aspect, for Taobao 2014, full-batch training achieves 0.0191 on recall@20 and for the best performing incremental learning approach, LWC-KD-PIW, achieves 0.0178 on recall@20. For Gowalla, full-batch training achieves 0.1963 on recall@20 and LWC-KD-PIW  achieves 0.2033 on recall@20, which is higher than the full-batch training paradigm. Thus, we can conclude that even on the model accuracy aspect, there is no guarantee that accumulating more training data from historic data will bring performance advantages.

\begin{figure}
    \centering
    \includegraphics[width=0.99\columnwidth]{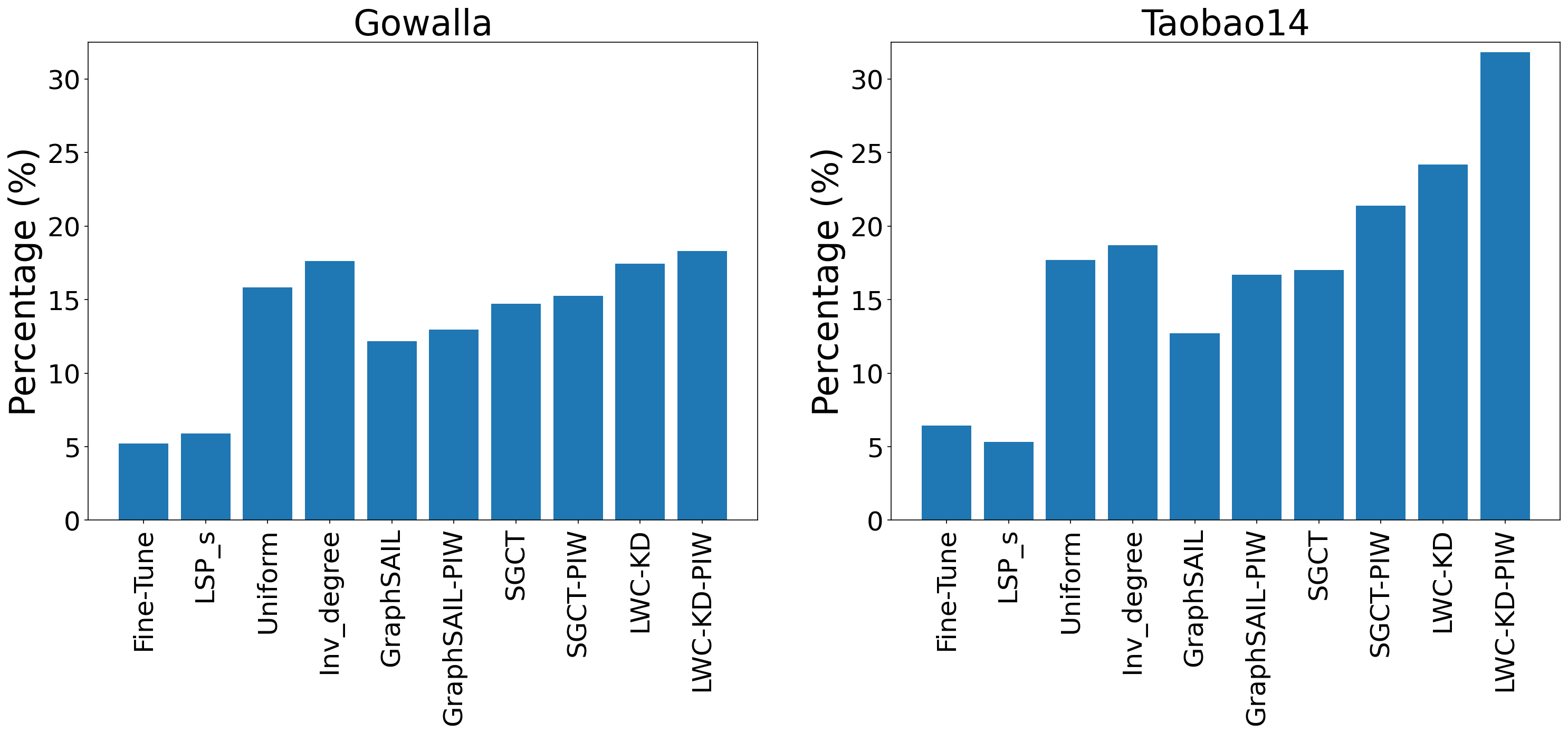}
    \caption{Percentage of time used compared to full-batch training. For Gowalla and Taobao2014 dataset, we have implemented a full-batch training version and then used its training time as a reference. We display the percentage of time used to finish training one incremental block compared to full-batch training time.}
    \label{fig:training time}
    \vspace{-1em}
\end{figure}

In a live deployment setting for a client-facing recommender system, the incremental training approach would be able to provide daily updates to the model, whereas full batch re-training would suffer from slow convergence  and unable to train the model with the update-to-date data stream,  leading to performance degradation ~\cite{xu2020graphsail}. We have compared the average training time of one incremental block t using incremental training-based methods for Taobao2014 and Gowalla to full batch training design (training with all blocks up to t and testing on t+1). As shown in Figure \ref{fig:training time}, even though LWC-KD-PIW  takes the longest time across all the incremental learning methods, it only takes 
at most 30\% of training time used by the full batch training design, which demonstrates that incremental learning has the potential to save significant training time.

\section{Concrete Realizations}
In this section, we provide three concrete realizations where we apply our adaptive imitation learning strategy in conjunction with three recent incremental learning techniques for recommender systems.

\paragraph{GraphSAIL~\cite{xu2020graphsail}} GraphSAIL (GS) implements distillation on each node's local neighborhood, global position relative to clusters of other nodes, and each node's own embedding. The distillation loss term $\mathcal{L}_{GS-adapt}$ (equation 12 in \cite{xu2020graphsail}) is modified to:
\begin{align}
\hspace{-2em}
\mathcal{L}_{GS-adapt} &= \frac{1}{|\mathcal{U}|} \sum_{u\in\mathcal{U}} {\color{blue}{w_u}} \bigg( \mathcal{L}_{self}(u)  + \mathcal{L}_{local}(u) + \mathcal{L}_{global}(u) \bigg) + \mathcal{L}_{KD}^{I},
\end{align}
where $\mathcal{U}$ is the set of user nodes and the GraphSAIL self, local and global loss components are defined in equations 3, 8, 9 in the GraphSAIL paper~\cite{xu2020graphsail}.
Here, $\textcolor{blue}{w_{u}}$ is the adaptive weight learned by the weight generating network.

\paragraph{SGCT \cite{Wang2021GraphSA}}
The contrastive distillation objective is used to push embeddings closer for positive pairs on latent space and push away negative pairs' embedding to distant embeddings. The positive pairs and negative pairs are constructed from user-item bipartite graph so to capture important structural information. The adaptive weights are then applied to users' contrastive distillation term so that we avoid ``pushing'' embeddings of users too close to previous positive neighbors if their interests have shifted a lot, i.e. they might no longer be interested in previous positive neighbors. Thus, the distillation objective term ($\mathcal{L}_{const}$ in \cite{Wang2021GraphSA}) becomes:
\begin{align}
\label{eqn:SGCT_adapt}
&\mathcal{L}_{sgct-adapt}  = \mathcal{L}_{KD}^{I} +  \frac{1}{|\mathcal{U}|}\sum_{u \in \mathcal{U}} \frac{-1}{|\mathcal{N}_{UI}^{t-1}(u)|} \nonumber\\ 
&\sum_{i \in \mathcal{N}_{UI}^{t-1}(u)} \textcolor{blue}{w_u}\log \frac{\exp \left(\boldsymbol{h}_{u,0}^{t} \cdot \boldsymbol{h}_{i,0}^{t-1} / \tau\right)}{\sum_{\hat{i} \in \mathcal{D}_{UI}^{t-1} (u)} \exp \left(\boldsymbol{h}_{u,0}^{t} \cdot \boldsymbol{h}_{\hat{i},0}^{t-1} / \tau\right)} \nonumber\\
\end{align}
where $\boldsymbol{h}_{u,0}^{t}$ is the embedding for node $u$ at time $t$, $\mathcal{N}_{UI}^{t-1}(u)$ is the neighborhood set of user node $u$ from the user-item interaction graph at time $t{-}1$, which provides the positive samples and $\mathcal{D}_{UI}^{t-1}(u)$ is the collection (union) of positive and negatives samples of the user $u$ generated from the user-item bipartite graph from time $t-1$. $\tau$ is a temperature that adjusts the concentration level.

\paragraph{LWC-KD~\cite{Wang2021GraphSA}}: On top of \textit{SGCT}, 
this method further takes advantage of (a) layer-wise distillation, which injects intermediate layer-level supervision; (b) contrastive distillation loss not only on user-item bipartite graph, but also on user-user and item-item graphs as well. The adaptive weights are applied to users' contrastive distillation terms based on user-item graph and user-user graph. Therefore, the model controls how much a user embedding is ``pushed'' to similar users or items of interest of previous time blocks based on how much his/her interest has shifted. The distillation objective ($\mathcal{L}_{lw-const}$ in \cite{Wang2021GraphSA}) is modified to: 
\begin{align}
\hspace{-2em} & \mathcal{L}_{lw-const-adapt} = \mathcal{L}_{KD}^{I} +  \frac{1}{K}\sum_{k=0}^K \frac{1}{|\mathcal{U}|}\sum_{u \in \mathcal{U}} \nonumber\\
\hspace{-2em} & \Bigg \{  \frac{-1}{|\mathcal{N}_{UI}^{t-1}(u)|} \sum_{i \in \mathcal{N}_{UI}^{t-1}(u)}  \textcolor{blue}{w_{u}}\log \frac{\exp \left(\boldsymbol{h}_{u,k}^{t} \cdot \boldsymbol{h}_{i,k}^{t-1} / \tau\right)}{\sum_{\hat{i} \in \mathcal{D}_{UI}^{t-1}(u)} \exp \left(\boldsymbol{h}_{u,k}^{t} \cdot \boldsymbol{h}_{\hat{i},k}^{t-1} / \tau\right)}  \nonumber\\
\hspace{-2em} & + \frac{-1}{|\mathcal{N}_{UU}^{t-1}(u)|} \sum_{u' \in \mathcal{N}_{UU}^{t-1}(u)} \textcolor{blue}{w_{u}}\log \frac{\exp \left(\boldsymbol{h}_{u,k}^{t} \cdot \boldsymbol{h}_{u',k}^{t-1} / \tau\right)}{\sum_{\hat{u} \in \mathcal{D}_{UU}^{t-1}} \exp \left(\boldsymbol{h}_{u,k}^{t} \cdot \boldsymbol{h}_{\hat{u},k}^{t-1} / \tau\right)}\Bigg \} 
\end{align}
where $\boldsymbol{h}_{u,k}^{t}$ is the embedding for node $u$ at time $t$ and layer $k$. $\mathcal{N}_{UU}^{t-1}(u)$ and $\mathcal{D}_{UU}^{t-1}(u)$ are the positive and negative neighborhood sets of user $u$ from the user-user similarity graph at time $t-1$, which provide samples for the contrastive objective of the corresponding term.

\section{Ablation studies}
To examine the effectiveness of each component --- weight\_generator, trans\_mat and cluster --- we additionally conduct experiments on the Gowalla and Taobao2015 datasets with the SGCT and LWC-KD algorithms. In each ablation, we remove one of components from our final method as shown in Table \ref{tab:method_components}. Therefore, the methods are formulated as:

\textbf{SAIL-PIW-no-wg}: No weight generator is used when learning weights. Therefore, after calculating eqn. \eqref{eqn: dist_G}, we directly calculate weights with $w_u =  ||\boldsymbol{G}_{u}^{t-1}-\boldsymbol{G}_{u}^{t}||^2$.

\begin{align}
      &\tilde{\boldsymbol{G}}_{u}^{t} = [\boldsymbol{\mu}_1^{t}\boldsymbol{W}_1 (\boldsymbol{h}_{u}^{t})^T,...,\boldsymbol{\mu}_M^{t}\boldsymbol{W}_M (\boldsymbol{h}_{u}^{t})^T]\,, \\
        & \boldsymbol{G}_{u,m}^{t} = \frac{e^{\tilde{\boldsymbol{G}}_{u,m}^{t}}}{\sum_{m'=1}^{M}e^{\tilde{\boldsymbol{G}}_{u,m'}^{t}}}\,,
\label{eqn: dist_G}
\end{align}

\textbf{SAIL-PIW-no-cluster}: Deep clustering is not applied to items when calculating the users' interest distribution over item clusters. Instead, we directly calculate the users' distribution change with respect to its item neighbors based on the user-item bipartite graphs of two consecutive blocks. Therefore, the state vector is designed as:
    \begin{align}
        \boldsymbol{d}_u^{t} &= [\boldsymbol{h}_{i_1}^{t}\boldsymbol{W} (\boldsymbol{h}_{u}^{t})^T,\dots,\boldsymbol{h}_{i_n}^{t}\boldsymbol{W} (\boldsymbol{h}_{u}^{t})^T] ~ \text{ where } ~ i_{1,\dots,n} \in \mathcal{N}_{UI}^{t-1}\,, \\
        \boldsymbol{s}_u &= ({d}_u^{t-1} - {d}_u^{t})\odot ({d}_u^{t-1} - {d}_u^{t})\,
    \end{align}
    where $\boldsymbol{W} \in \mathcal{R}^{d \times d}$ is a transformation matrix and the same item set is used for both time blocks. We obtain $\boldsymbol{d}_u^{t-1}$ similarly to $\boldsymbol{d}_u^{t}$.

\textbf{SAIL-PIW-no-trans}: No transformation is used in this design. Thus, the step in eqn. \ref{eqn: dist_G} becomes 
\begin{equation}
    \tilde{G}_{u}^{t} = [\boldsymbol{\mu}_1^{t} (\boldsymbol{h}_{u}^{t})^T,...,\boldsymbol{\mu}_M^{t} (\boldsymbol{h}_{u}^{t-1})^T]
\end{equation}

\textbf{SAIL-PIW-hard}: In addition to the previous designs, we also do one more ablation study where K-means is used instead of the deep clustering network to obtain the average embeddings of the clusters $\boldsymbol{\mu}_m^{t}$. To be specific, we apply K-means on $\boldsymbol{h_{i}^{t-1}}$ and identify M clusters. Then we set $\boldsymbol{\mu}_m^{t} = \frac{1}{|C_{m}^{t}|}\sum_{i \in C_{m}^{t-1}}(\boldsymbol{h}_{i}^{t})$, where $C_{m}^{t-1}$ denotes the item set of cluster $m$.

We observe in Table~\ref{tab:ablation} that the removal of any component leads to performance deterioration, as does the absence of end-to-end training for clustering. Therefore, we conclude that each aspect of the design of the algorithm contributes to its outperformance.
We also evaluate how the model performs as we vary the number of clusters using SGCT and LWC-KD in Table \ref{tab:ablation_K}. We observe that the performance does not depend strongly on the choice of K. In other words, the model is robust to the choice of the number of clusters. There is some evidence of a slight drop in overall performance if we choose a small or large number of clusters.
\begin{table}[ht]
\caption{Method components for ablation analysis.}
\vspace{-0.3cm}
  \label{tab:method_components}
  \resizebox{\columnwidth}{!}{\begin{tabular}{c|ccc}
\toprule
Method & weight\_generator &trans\_mat &cluster\\
SAIL-PIW-no-wg &\xmark&\checkmark&\checkmark\\
SAIL-PIW-no-cluster &\checkmark&\checkmark&\xmark\\
SAIL-PIW-no-trans &\checkmark&\xmark&\checkmark\\
SAIL-PIW-hard&\checkmark&\checkmark&\checkmark\\
SAIL-PIW&\checkmark&\checkmark&\checkmark\\
\bottomrule
\end{tabular}}
 \vspace{-0.1cm}
\end{table}
\begin{table}[ht]
    \centering
    \caption{Ablation analysis on the state vector design. Recall@20 results on the Gowalla and Taobao2015 datasets for the SGCT and LWC-KD distilling algorithms.}
    \label{tab:ablation}
    \resizebox{\columnwidth}{!}{\begin{tabular}{c|c|c|cccc}
    \toprule
  Distillation Strategies & Dataset & Algorithm & Inc 1 & Inc 2 & Inc 3 & Avg. Recall@20\\
    \midrule
    \multirow{10}{*}{SGCT}&  \multirow{5}{*}{Gowalla} & SAIL-PIW-no-wg & 0.1535 & 0.1807 & 0.2221 & 0.1854\\
    & & SW-AIW-click-uni-interest & 0.1525 & 0.1828 & 0.2169 & 0.1841\\
    & &SAIL-PIW-no-trans & 0.1497 & 0.1842 & 0.2225 & 0.1855\\
    & & SAIL-PIW-hard & 0.1573 & 0.1860 & 0.2212 & 0.1882\\
    & &SAIL-PIW &0.1599&0.1892&0.2321&0.1937\\
   \cline{2-7}

    & \multirow{5}{*}{Taobao2015} & SAIL-PIW-no-wg &0.1022&0.0998&0.1013&0.1011\\
    & & SW-AIW-click-uni-interest &0.1019&0.0992&0.1012&0.1008\\
    & &SAIL-PIW-no-trans &0.1021&0.0987&0.1000&0.1003\\
    & & SAIL-PIW-hard &0.0972&0.0975&0.0989&0.0979\\
    & &SAIL-PIW  &0.1040&0.0999&0.1027&0.1022\\
    \midrule
    \multirow{10}{*}{LWC-KD} & \multirow{5}{*}{Gowalla} & SAIL-PIW-no-wg &0.1561&0.1861&0.2350&0.1924\\
    & & SW-AIW-click-uni-interest  &0.1596&0.1865&0.2337&0.1933\\
    & &SAIL-PIW-no-trans  &0.1631&0.1875&0.2288&0.1931\\
    & & SAIL-PIW-hard  &0.1604&0.1889&0.2246&0.1913\\
    & &SAIL-PIW  &0.1698&0.1978&0.2425&0.2033\\
\cline{2-7}

    & \multirow{5}{*}{Taobao2015} & SAIL-PIW-no-wg  &0.1007&0.1000&0.1032&0.1013\\
    & & SW-AIW-click-uni-interest  &0.1022&0.1006&0.1021&0.1016\\
    & &SAIL-PIW-no-trans  &0.1004&0.1007&0.1035&0.1015\\
    & & SAIL-PIW-hard  &0.1032&0.1012&0.1031&0.1025\\
    & &SAIL-PIW   &0.1044&0.1045&0.1052&0.1047\\
    \bottomrule
    \end{tabular}}
\end{table}

\begin{table}[ht]
    \centering
    \caption{Ablation analysis for the cluster number K. Recall@20 results on the Gowalla and Taobao2015 datasets for the SGCT and LWC-KD distilling algorithms.}
    \label{tab:ablation_K}
    \resizebox{\columnwidth}{!}{\begin{tabular}{c|c|c|cccc}
    \toprule
  Distillation Strategies & Dataset & K & Inc 1 & Inc 2 & Inc 3 & Avg. Recall@20\\
    \midrule
    \multirow{10}{*}{SGCT}&  \multirow{5}{*}{Gowalla} & 5 & 0.1608 & 0.1846 & 0.2192 & 0.1882\\
    & & 10 &0.1599&0.1892&0.2321&0.1937\\
    & & 15 & 0.1601 & 0.1956 & 0.2244 & 0.1934\\
    & & 20 & 0.1557 & 0.1904 & 0.2284 & 0.1915\\
    & & 25 & 0.1549 & 0.1918 & 0.2259 &0.1909\\
   \cline{2-7}

    & \multirow{5}{*}{Taobao2015} & 5 0&0.0981&0.1007&0.1014&0.1001\\
    & & 10 &0.1040&0.0999&0.1027&0.1022\\
    & & 15&0.0996&0.1000&0.1040&0.1012\\
    & & 20&0.1025&0.1005&0.1032&0.1021\\
    & & 25 &0.1018&0.0983&0.1016&0.1006\\
    \midrule
    \multirow{10}{*}{LWC-KD} & \multirow{5}{*}{Gowalla} & 5 &0.1727&0.2028&0.2346&0.2034\\
    & & 10  &0.1698&0.1978&0.2425&0.2033\\
    & &15  &0.1663&0.1945&0.0.2411&0.2006\\
    & & 20  &0.1704&0.1992&0.2399&0.2032\\
    & &25 &0.1707 &0.1950&0.2371&0.2009\\
\cline{2-7}

    & \multirow{5}{*}{Taobao2015} & 5 &0.0993&0.1014&0.1043&0.1017\\
    & & 10  &0.1044&0.1045&0.1052&0.1047\\
    & & 15 &0.1045&0.1039&0.1039&0.1041\\
    & & 20 &0.1017&0.1028&0.1025&0.1023\\
    & & 25  &0.1017&0.1028&0.1041&0.1029\\
    \bottomrule
    \end{tabular}}
\end{table}

\section{Case Study - Additional Details}

We have conducted a case study in order to more closely examine how three models behave for two distinct groups of users. The three models we study are fine-tune, GraphSAIL without adaptive weights and GraphSAIL with adaptive weights. We identify two types of users: (i) 
{\em static} users who exhibit minimal interest shift; and {\em dynamic} users who exhibit dramatic interest shift. Then we test the models on the historical data (i.e., data from previous time block). The detailed steps of the case study on how we determine user shift are:
\begin{enumerate}
    \item Perform K means on item embeddings $\boldsymbol{h}_i^{t-1}$ obtained from the GraphSAIL model to identify $M$ clusters. Here $t$ is the training block number.
    \item Count the number of items each user interacted with in each cluster to construct the user's interest histogram for both time block $t$ and block $t-1$. We calculate $\Tilde{\boldsymbol{I}}^{t} \in \mathbb{R}^{U \times M}$ and $\Tilde{\boldsymbol{I}}^{t-1} \in \mathbb{R}^{U \times M}$.
    \item Normalize $\Tilde{\boldsymbol{I}}$ to calculate $ \boldsymbol{I}_m= \Tilde{\boldsymbol{I}}_m/\sum_{m' \in M}\Tilde{\boldsymbol{I}}_m'$. 
    \item Calculate users' interest shift indicator score:\\ $ISS_u = \frac{1}{M}\sum_M||\boldsymbol{I}^{t}_u-\boldsymbol{I}^{t-1}_u||^2$.
    
    \item Select users with top 20\% interest shift indicator scores to form the {\em dynamic} group and users with bottom 20\% Interest shift indicator scores to form the {\em static} group and record the corresponding user indices.
    \item Calculate separately for the two groups of users the average recall for fine-tune, GraphSAIL and GraphSAIL-PIW.
\end{enumerate}

\bibliography{main.bib}